\documentclass[structabstract]{aa}
\usepackage{graphicx}
\usepackage{txfonts}
\usepackage{natbib}
\usepackage{longtable}

\bibpunct[]{(}{)}{;}{a}{}{,}
\begin{document}
   \title{Rotational spectra of isotopic species of methyl cyanide, CH$_3$CN, in 
          their ground vibrational states up to terahertz frequencies}

   \author{Holger~S.~P. M{\"u}ller\inst{1}
           \and
           Brian~J. Drouin\inst{2}
           \and
           John~C. Pearson\inst{2}
          }

   \institute{I.~Physikalisches Institut, Universit{\"a}t zu K{\"o}ln,
              Z{\"u}lpicher Str. 77, 50937 K{\"o}ln, Germany\\
              \email{hspm@ph1.uni-koeln.de}
         \and
              Jet Propulsion Laboratory, California Institute of Technology, 
              Mail Stop 183-301, 4800 Oak Grove Drive, 
              Pasadena, CA 91011-8099, USA}

   \date{Received 21 July 2009 / Accepted 18 August 2009}

  \abstract
{Methyl cyanide is an important trace molecule in star-forming regions. 
It is one of the more common molecules used to derive kinetic temperatures 
in such sources.}
{As preparatory work for Herschel, SOFIA, and in particular ALMA we want to improve 
the rest frequencies of the main as well as minor isotopologs of methyl cyanide.}
{The laboratory rotational spectrum of methyl cyanide in natural isotopic 
composition has been recorded up to 1.63~THz.}
{Transitions with good signal-to-noise ratio could be identified for 
CH$_3$CN, $^{13}$CH$_3$CN, CH$_3^{13}$CN, CH$_3$C$^{15}$N, 
CH$_2$DCN, and $^{13}$CH$_3^{13}$CN in their ground vibrational states 
up to about 1.2~THz. 
The main isotopic species could be identified even in the highest frequency 
spectral recordings around 1.6~THz. 
The highest $J'$ quantum numbers included in the fit are 64 
for $^{13}$CH$_3^{13}$CN and 89 for the main isotopic species.
Greatly improved spectroscopic parameters have been obtained by fitting 
the present data together with previously reported transition frequencies.}
{The present data will be helpful to identify isotopologs of 
methyl cyanide in the higher frequency bands of instruments 
such as the recently launched Herschel satellite, the upcoming 
airplane mission SOFIA or the radio telescope array ALMA.}

\titlerunning{rotational spectra of CH$_3$CN isotopic species}

\maketitle
\hyphenation{cyano-methane}
\hyphenation{Na-ti-o-nal}
\hyphenation{con-tri-bu-tion}
\hyphenation{Sym-po-si-um}
\hyphenation{Co-lum-bus}

%

\section{Introduction}
\label{intro}

Methyl cyanide, CH$_3$CN, also known as acetonitrile, or cyanomethane, 
was among the early molecules to be detected by radio astronomical means. 
\citet{det-MeCN} detected it almost 40 years ago toward the massive 
star-forming regions Sagittarius A and B close to the Galactic center. 
It has been detected in dark clouds such as TMC-1 \citep{MeCN-TMC-1}, 
around the low-mass protostar IRAS 16293$-$2422 \citep{MeCN_IRAS16293}, 
in the circumstellar shell of the famous carbon-rich star IRC~+10216 
\citep{MeCN_etc_IRC+10216}, in comets, such as Kohoutek 
\citep{MeCN-Kohoutek}, and in external galaxies \citep{MeCN_MeCCH_extragal}.
As a strongly prolate symmetric top, meaning $A \gg B$, transitions 
with different $K$ but the same $J$ occur in fairly narrow 
frequency regions, but sample rather different energies. 
Therefore, CH$_3$CN is quite commonly used to derive the kinetic 
temperatures of dense molecular clouds as already done in \citet{det-MeCN} 
or in e.\,g. \citet{MeCN-T_kin}. In less dense clouds the large 
dipole moment of 3.922~D \citep{MeCN-dipole} may prevent 
thermal equilibration. In that case, the isoelectronic 
methyl acetylene, or propyne, CH$_3$CCH, is often used instead. 

The $^{12}$C/$^{13}$C ratio in the vicinity of the Galactic center is 
approximately 20 \citep{isotopic_abundances,13C-VyCN_2008}. 
Thus, it is not surprising that the two isotopomers containing 
$^{13}$C were detected fairly soon after the detection of 
the main isotopolog. CH$_3 ^{13}$CN \citep{MeCN-T_kin} was 
detected first probably by accident because the isotopic substitution 
at the central C-atom does not change the spectroscopic parameters 
much with respect to the main species. \citet{Orion-survey_13CH3CN} 
detected lines of both isotopomers with one $^{13}$C in a line survey 
of Orion. More recently, CH$_2$DCN has been detected in two hot 
core sources \citep{det-CH2DCN}, and some lines of CH$_3$C$^{15}$N 
were detected in a line survey of Sagittarius~B2 \citep{Sgr-B2_Nummelin}.

A plethora of high-resolution spectroscopic investigations have been 
performed on methyl cyanide. It was actually among the first molecules 
to be studied by microwave spectroscopy by \citet{MeCN_1st-MW}. 
\citet{MeCN-LD} reviewed the investigations for the main isotopic species 
and performed high resolution, high accuracy measurements up to 1.2~THz. 
\citet{MeCN_rot-isos_1996} analogously presented data up to 607~GHz 
for the singly substituted $^{13}$C isotopomers as well as for 
the species with $^{15}$N. \citet{CH2DCN-rot} built on the very small 
data set for CH$_2$DCN by measuring an isotopically enriched sample 
up to 471~GHz. Finally, there has been only one report on the isotopolog 
with two $^{13}$C which was restricted to frequencies up to 72~GHz 
\citep{MeCN-2x13}.

We have measured rotational spectra of methyl cyanide in natural isotopic 
composition both in wide frequency windows as well as selected lines 
up to 1.63~THz to provide new or updated catalog entries for various 
isotopologs of methyl cyanide as well as for excited vibrational states. 
With rotational temperatures of CH$_3$CN in hot cores reaching at least 
150$-$200~K \citep{Orion-survey_13CH3CN,det-PrCN_EtFo}, these data 
will be of great relevance for the Atacama Large Millimeter Array (ALMA). 
The main isotopolog and possibly even the $^{13}$C isotopomers or the 
$\varv _8 = 1$ excited vibrational state will be seen with the 
HIFI instrument (Heterodyne Instrument for the Far-Infrared) on board 
of the recently launched Herschel satellite or with SOFIA (Stratospheric 
Observatory For Infrared Astronomy). 
In the present article we provide the ground state rotational data 
obtained for six different isotopic species: CH$_3$CN, $^{13}$CH$_3$CN, 
CH$_3^{13}$CN, CH$_3$C$^{15}$N, CH$_2$DCN, and $^{13}$CH$_3^{13}$CN; 
as usual, unlabeled atoms refer to $^{12}$C and $^{14}$N. 
Analyses of selected excited vibrational states are currently 
under way.

\section{Experimental details}
\label{exptl}

Individual transitions have been recorded with the Cologne Terahertz 
Spectrometer (CTS) which has been described in detail by 
\citet{BWO-THz_spec}. It uses broadband tunable, phase-locked
backward-wave oscillators (BWOs) as powerful sources, and a magnetically
tuned, liquid helium cooled hot-electron InSb bolometer as detector.
In the present case, one BWO was used to record transitions between 
249 and 340~GHz.

The pressure of methyl cyanide in the 3~m long absorption cell 
was around 0.3$-$0.5~Pa. The measurements were carried out in static
mode, and the sample was pumped off and replaced about every half hour.

The accuracies with which lines can be measured with the CTS depend 
on the lines shape, mostly on how symmetric the line is, and on the 
signal-to-noise (S/N) ratio. 
Quite commonly, we reach relative accuracies of $10^{-8}$ or 
even slightly better \citep{SO2_2005,SiS_2007,13C-VyCN_2008}.
\citet{SO-17O} have shown that similar accuracies can be achieved 
for the rare isotopolog SO$^{17}$O in a rather dense spectrum. 
The lines recorded for $^{13}$CH$_3$CN, CH$_3^{13}$CN, and CH$_3$C$^{15}$N 
had very good S/N ratios and line shapes. Thus, the uncertainties 
were judged to be $\le 10$~kHz. The lines of CH$_2$DCN and 
$^{13}$CH$_3^{13}$CN were considerably weaker such that uncertainties 
of up to 30~kHz were assigned.

The majority of the data has been extracted from broad frequency scans 
taken with the JPL cascaded multiplier spectrometer \citep{JPL-spectrometer}. 
Generally, a multiplier chain source is passed through a $1-2$ meter pathlength 
flow cell and is detected by a silicon bolometer cooled to near 1.7~K. 
The cell is filled with a steady flow of reagent grade acetonitrile 
and the pressure and modulation are optimized to enable good S/N ratios 
with narrow lineshapes.
With a gas with very strong transitions, such as the $K < 7$ transitions 
of the main isotopolog of acetonitrile, the S/N ratio was optimized for 
a higher-$K$ transition (e.g. $K = 12$) such that the lower $K$ transitions 
exhibit saturated line profiles. 
This procedure enables better dynamic range for the extraction of line positions 
for rare isotopologs and highly excited vibrational satellites. 
The frequency ranges covered were 440$-$540, 619$-$631, 638$-$648, 
770$-$855, 875$-$930, 967$-$1050, 1083$-$1093, 1100$-$1159, 
1168$-$1198, 1576$-$1591, and 1614$-$1627~GHz. Most of these multiplier sources 
were previously described \citep{JPL-spectrometer}. 
However, the multiplier chain with frequency range coverage between 
967$-$1050~GHz was not described in that work. 
This chain consists of two cascaded triplers after the amplified W-band stage, 
the peak output power is near 100~$\mu$W. 
The efficiency of frequency multipliers usually changes strongly with frequency. 
In addition, recording conditions and sensitivities of detectors can have 
strong influences on the quality of the spectra. 
Particularly good S/N ratios were reached around 630, 900 and around 
1100$-$1200~GHz. The spectra around 500 and around 800~GHz had poorer 
S/N ratios. In addition, the S/N ratios changed considerably 
within each scan and were usually lower towards the ends. 
Uncertainties of 50~kHz were assigned to isolated lines with good to 
very good S/N ratios throughout the frequency regions. 
Larger uncertainties were assigned to weaker lines or lines 
which were not isolated. In the course of the analysis it was observed 
that several strong lines had residuals considerably smaller than 50~kHz. 
Thus, smaller uncertainties, down to 20~kHz, were assigned to 
very strong lines, depending on the symmetry of the line shape.


\begin{table}
\begin{center}
\caption{Lower state quantum numbers of rotational transitions$^a$ 
of CH$_3$CN from present work, frequencies (MHz), uncertainties unc. (kHz), and 
residuals o$-$c (kHz) between observed frequencies and those calculated from 
the final set of spectroscopic parameters.}
\label{tab:MeCN}
\begin{tabular}[t]{rrr@{}lrr}
\hline \hline
$J''$ & $K''$ & \multicolumn{2}{c}{frequency} & 
unc. & o$-$c \\
\hline
 86 & 15 &  1584230&.365 & 150 & $-$155 \\
 86 & 14 &  1585042&.686 & 100 &     39 \\
 86 & 13 &  1585800&.110 & 100 &      4 \\
 86 & 12 &  1586502&.584 &  50 &  $-$32 \\
 86 & 11 &  1587149&.953 &  50 &     35 \\
 86 & 10 &  1587741&.807 &  50 &     39 \\
 86 &  9 &  1588277&.964 &  50 &     22 \\
 86 &  8 &  1588758&.221 &  50 &  $-$15 \\
 86 &  7 &  1589182&.509 &  50 &     44 \\
 86 &  6 &  1589550&.454 &  50 &  $-$11 \\
 86 &  5 &  1589862&.092 &  50 &   $-$2 \\
 86 &  4 &  1590117&.262 &  50 &     32 \\
 86 &  3 &  1590315&.822 &  50 &     50 \\
 86 &  2 &  1590457&.643 &  50 &   $-$1 \\
 86 &  1 &  1590542&.824 &  50 &     34 \\
 86 &  0 &  1590571&.147 &  50 &  $-$28 \\
 87 &  9 &  1606303&.437 & 100 &     75 \\
 87 &  8 &  1606788&.347 &  50 &     71 \\
 87 &  7 &  1607216&.562 &  50 &  $-$25 \\
 87 &  6 &  1607588&.178 &  50 &     50 \\
 87 &  5 &  1607902&.790 &  50 &     35 \\
 87 &  4 &  1608160&.351 &  50 &      6 \\
 87 &  3 &  1608360&.834 &  50 &     36 \\
 87 &  2 &  1608504&.060 &  50 &     25 \\
 87 &  1 &  1608590&.002 &  50 &      3 \\
 87 &  0 &  1608618&.661 &  50 &      3 \\
 88 & 10 &  1623774&.391 &  50 &      9 \\
 88 &  9 &  1624320&.812 &  50 &  $-$33 \\
 88 &  8 &  1624810&.328 &  50 &  $-$27 \\
 88 &  7 &  1625242&.737 &  50 &     13 \\
 88 &  6 &  1625617&.728 &  50 &  $-$56 \\
 88 &  5 &  1625935&.334 &  50 &  $-$57 \\
 88 &  4 &  1626195&.369 &  50 &  $-$52 \\
 88 &  3 &  1626397&.725 &  50 &  $-$47 \\
 88 &  2 &  1626542&.355 &  50 &  $-$11 \\
 88 &  1 &  1626629&.131 &  50 &  $-$14 \\
 88 &  0 &  1626658&.067 &  50 &   $-$7 \\
\hline
\end{tabular}\\[2pt]
\end{center}
$^a$ Transitions are $J'' +1, K'' \leftarrow J'', K''$ for all isotopologs 
except for CH$_2$DCN.
\end{table}

\section{Analysis and discussion}
\label{a_and_d}

The previously available data set will be described in the following 
for each isotopolog individually, with the exception of the two singly 
substituted $^{13}$C species because the latter had almost always been 
investigated together. The extent of lines added in the course of the 
present work will be given also.

The main isotopolog of methyl cyanide as well as all other ones 
investigated in the present work are strongly prolate symmetric rotors 
having $C_{\rm 3v}$ symmetry, except for CH$_2$DCN. 
Spin-statistics have to be taken into account for the 
symmetric rotor isotopologs. Levels having $K = 3n \pm 1$ 
belong to the $E$ symmetry class whereas levels having $K = 3n$ 
belong to the $A$ symmetry class; $n \ge 0$ in all instances. 
The spin-statistical weight of $A$ symmetry levels with $K > 0$ 
is twice that of $K = 0$ and all $E$ symmetry levels. 

The observed transitions all obey the $\Delta K = 0$ selection rule 
thus their positions are unaffected by the purely $K$-dependent parameters. 
However, these parameters are not negligible as they affect 
the intensities of the $\Delta K = 0$ transitions due 
to their contribution to the energies of the rotational levels.

In a strongly prolate molecule, such as methyl cyanide, $\Delta K = 3$ 
transitions only gain intensity through centrifugal distortion effects; 
they were too weak to be identified unambigously in the present work. 
Because of this, the purely $K$-dependent parameters $A$, $D_K$, etc. 
are usually not determinable for such molecules by means of 
rotational spectroscopy. 

The dipole moment has been measured both for CH$_3$CN \citep{MeCN-dipole}
and CH$_3$C$^{15}$N \citep{MeCN-15-dipole}. Isotopic differences in the 
dipole moments are expected to be small for the symmetric top species. 
In the case of CH$_3$C$^{15}$N, a value was obtained that deviated 
barely significantly (3.9256~(7)~D) from the value for the 
main isotopolog (3.92197~(13)~D). 

CH$_2$DCN is the only asymmetric top rotor among the isotopologs 
dealt with in the present article. It only has $C_S$ symmetry, 
therefore, no spin-statistics have to be considered, in contrast 
to what was assumed in \citet{det-CH2DCN}. 
The asymmetry parameter $\kappa = (2B - A - C)/(A - C)$ is $-$0.9973, 
expectedly close to the prolate symmetric top limit of $-1$. 
The small rotation of the inertial axis system caused 
by the substitution of one H atom by D gives rise to a small 
$b$-dipole moment component of $\sim$0.17~D. 
$\Delta K_a$ and $\Delta K_c$ have to be odd for $b$-type  transitions. 
However, most of the previously observed and all of the new transitions 
are $a$-type transitions with $\Delta K_a = 0$ and $\Delta K_c = 1$. 
Transitions having $\Delta K_a = 2$ are allowed also, but are very weak 
for asymmetric molecules close to the prolate symmetric limit.


\begin{table*}
\begin{center}
\caption{Spectroscopic parameters$^a$ (MHz) of methyl cyanide isotopic 
species with $C_{\rm 3v}$ symmetry and dimensionless weighted standard 
deviation wrms.}
\label{parameter_C3v}
\renewcommand{\arraystretch}{1.10}
\begin{tabular}[t]{lr@{}lr@{}lr@{}lr@{}lr@{}l}
\hline \hline
Parameter & \multicolumn{2}{c}{CH$_3$CN} & \multicolumn{2}{c}{CH$_3^{13}$CN} &
 \multicolumn{2}{c}{$^{13}$CH$_3$CN}  & \multicolumn{2}{c}{$^{13}$CH$_3^{13}$CN} & \multicolumn{2}{c}{CH$_3$C$^{15}$N} \\
\hline
$(A - B) \times 10^{-3}$        &    148&.900\,074\,(65) &    148&.904\,624       &    149&.165\,664       &    149&.171\,692        &    149&.176\,934       \\
$B$                             & 9\,198&.899\,134\,(11) & 9\,194&.349\,983\,(27) & 8\,933&.309\,412\,(28) & 8\,927&.281\,294\,(155) & 8\,922&.038\,611\,(72) \\
$D_K$                           &      2&.825\,1\,(15)   &      2&.825\,7         &      2&.825\,7         &      2&.825\,7          &      2&.825\,7         \\
$D_{JK} \times 10^3$            &    177&.407\,96\,(28)  &    176&.673\,95\,(132) &    168&.239\,65\,(130) &    167&.420\,4\,(177)   &    168&.938\,0\,(34)   \\
$D_J \times 10^3$               &      3&.807\,528\,(9)  &      3&.809\,706\,(37) &      3&.624\,918\,(32) &      3&.627\,152\,(61)  &      3&.555\,206\,(45) \\
$H_K \times 10^6$               &     51&.               &     51&.               &     51&.               &     51&.                &     51&.               \\
$H_{KJ} \times 10^6$            &      6&.063\,1\,(16)   &      6&.001\,9\,(131)  &      5&.803\,1\,(141)  &      5&.50\,(42)        &      5&.641\,0\,(181)  \\
$H_{JK} \times 10^6$            &      1&.025\,61\,(18)  &      1&.017\,52\,(86) &      0&.927\,05\,(86)  &      0&.915\,9\,(35)    &      0&.950\,83\,(131) \\
$H_J \times 10^{12}$            & $-$255&.96\,(245)      & $-$270&.2\,(61)        & $-$283&.0\,(49)        & $-$298&.                & $-$197&.7\,(67)        \\
$L_{KKJ} \times 10^{12}$        & $-$446&.3\,(26)        & $-$440&.               & $-$433&.               & $-$427&.                & $-$425&.               \\
$L_{JK} \times 10^{12}$         &  $-$52&.58\,(64)       &  $-$49&.40\,(236)      &  $-$49&.45\,(266)      &  $-$47&.3               &  $-$48&.               \\
$L_{JJK} \times 10^{12}$        &   $-$7&.889\,(45)      &   $-$7&.730\,(141)     &   $-$6&.843\,(128)     &   $-$6&.7               &   $-$7&.041\,(182)     \\
$L_J \times 10^{15}$            &   $-$1&.80\,(19)       &   $-$1&.80             &   $-$1&.80             &   $-$1&.80              &   $-$1&.28             \\
$P_{JK} \times 10^{18}$         &    529&.\,(119)        &    530&.               &    450&.               &    450&.                &    440&.               \\
$P_{JJK} \times 10^{18}$        &     56&.1\,(29)        &     56&.               &     48&.               &     48&.                &     49&.               \\
$eQq$                           &   $-$4&.223\,24\,(108) &   $-$4&.218\,30\,(176) &   $-$4&.218\,13\,(197) &   $-$4&.213\,19         &       &n.\,a.          \\
$C_{bb} \times 10^3$            &      1&.785\,(92)      &      1&.785            &      1&.73             &      1&.73              &       &$-$             \\
$(C_{aa} - C_{bb}) \times 10^3$ &   $-$1&.24\,(31)       &   $-$1&.24             &   $-$1&.24             &   $-$1&.24              &       &$-$             \\
wrms                            &      0&.906            &      0&.813            &      0&.760            &      0&.635             &      0&.692            \\
\hline
\end{tabular}\\[2pt]
\end{center}
$^a$ Numbers in parentheses are one standard deviation in units of the least significant figures. 
Parameters without quoted uncertainties have been estimated from the main isotopic species and were kept fixed in the fits; 
see section~\ref{a_and_d}.\\
\end{table*}


\subsection{CH$_3$CN}
\label{main-iso}

\citet{MeCN-LD} carried out extensive high resolution, high accuracy 
measurements on CH$_3$CN up to 1.2~THz. Their data set also included 
data from \citet{MeCN_1-0,MeCN-12-13b_2-1,MeCN-Lille} below 74~GHz.
\citet{MeCN-v08} pointed to an interaction between $\varv = 0$, $K = 14$ 
and the lowest excited, doubly degenerate $\varv _8 = 1$ bending mode 
at $K = 12$ in the lower energy $l = +1$ component causing non-negligible 
shifts between $J'' \approx 36$ and 48. However, they were only able 
to perform a preliminary analysis because they only observed $J'' = 38$ 
and 39 of $\varv = 0$. A more complete analysis has been presented by 
\citet{MeCN-OSU}; a manuscript on this work, involving extensive 
rotational and rovibrational data for states with $\varv _8 \le 2$, 
is in preparation. The perturbation is negligible for all but the small number 
of $\varv = 0$, $K = 14$ transitions. Thus, in the present analysis, 
the perturbed transitions were omitted from the final fit. 
No evidence for perturbations was found in the 
rotational spectra of the less abundant isotopologs.

37 lines have been detected around 1.6~THz covering $J'' = 86$ to 88 
with $K$ up to 15. These data are given in Table~\ref{tab:MeCN}
together with their assignments, uncertainties, and the residuals 
between measured frequencies and those calculated from the 
final spectroscopic parameters. 
They have been fit together with previous data mentioned above 
which extended to $K = 21$ and $J'' = 64$. 
Also included in the fits were the $\Delta K = 3$ 
ground state energy differences from \citet{MeCN-A} 
in order to determine the purely $K$-dependent terms $A$ and $D_K$. 
Tunable far-infrared data from \citet{MeCN-TuFIR} were omitted from 
the final fits because trial fits with these data included showed 
their effect on the spectroscopic parameters to be negligible.

The considerable extension of the data set in $J$ required two additional 
higher order parameters to be included in the fit, the decic centrifugal 
distortion parameters $P_{JK}$ and $P_{JJK}$, as shown in 
Table~\ref{parameter_C3v}. On the other hand, the centrifugal 
distortion correction term $eQq_J$ has been omitted from the final 
fit because it was not significantly determined, $(31 \pm 15)$~Hz, 
and because it may still be too large in magnitude even though the 
absolute value in \citet{MeCN-LD} was more than one order of 
magnitude smaller than that in \citet{MeCN-v08}. 
If one assumes the ratio of $-D_J/B$ to be a good estimate 
for the ratio of $eQq_J/eQq$ then the $eQq_J$ value 
in \citet{MeCN-LD} is about one order of magnitude too large, 
but has presumably the correct (positive) sign; \citet{MeCN-v08} 
determined a negative value. As the hyperfine parameters 
and the lower order distortion terms are mostly determined by 
the data published by \citet{MeCN-LD} it is not surprising that 
there is a good agreement between the parameter values and uncertainties 
determined both in that study as well as in the present work. 
The increase in $J$ in the present study causes the purely $J$-dependent 
terms ($B$, ..., $L_J$) to be improved in accuracy 
despite the additional centrifugal distortion terms. 

The purely $K$-dependent terms $A$ and $D_K$ are determined in the 
present study entirely by the $\Delta K = 3$ ground state energy 
differences from \citet{MeCN-A}. Thus, good agreement is not surprising. 
Trial fits with $H_K$ released yielded a value of ($155 \pm 102)$~Hz 
as in \citet{MeCN-A}. 
Judging by the $D_K/A$ ratio, it is too large by about a factor of 3. 
Moreover, the parameter has not been determined with significance. 
Therefore, $H_K$ was fixed in the present analysis to a value 
derived from the $D_K/A$ ratio. Current results from the more extended 
analysis \citep{MeCN-OSU} are in support of this value.

The weighted standard deviation of the fit is 0.906, indicating that the 
data have been reproduced within experimental uncertainties on average. 
The corresponding value for only the new transition frequencies is 0.670, 
suggesting the estimates of the uncertainties to be slightly too 
conservative.


\begin{figure}
\includegraphics[angle=0,width=9cm]{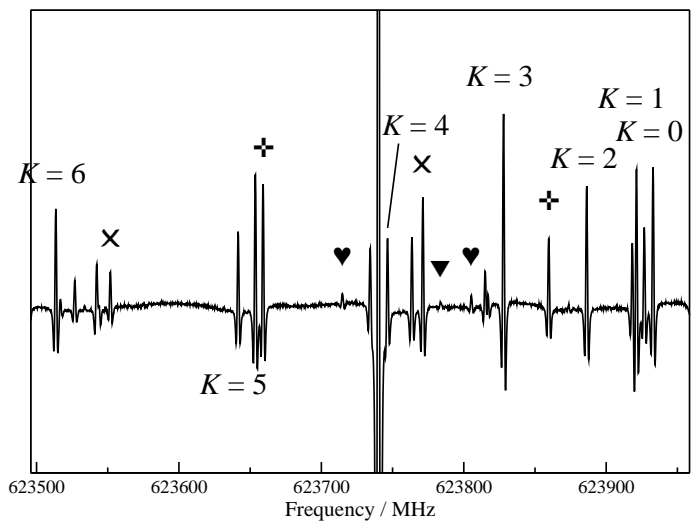}
  \caption{Section of the submillimeter spectrum of CH$_3$CN. 
   Transitions of CH$_3$C$^{15}$N have been labeled with their $K$ 
   quantum numbers. $K = 3$ and 6 appear stronger than expected because 
   of the spin-statistics, see section~\ref{a_and_d}. Also labeled are 
   $K = 10$ and 9 of $^{13}$CH$_3$CN by x-signs, $K = 9$ and 8 of 
   CH$_3^{13}$CN by plus-signs, $K = 9$ of CH$_2$DCN by an inverted 
   triangle, and $k = -15$ and +17 of CH$_3$CN, $\varv _8 = 1$ by 
   heart-signs. The strong, clipped line in the center of the figure 
   is due to $K = 10$ of CH$_3$CN, $\varv = 0$. The remaining 
   unlabeled transitions have not been assigned thus far.
   $K = 10$ of CH$_2$DCN as well as three transitions of 
   $^{13}$CH$_3^{13}$CN are too weak to be recognized on this scale. 
   The lines appear as second 
   derivatives of a Gaussian line-shape because of the 2$f$-modulation.}
\label{isos}
\end{figure}



\subsection{$^{13}$CH$_3$CN and CH$_3^{13}$CN}
\label{13C}

The data set of \citet{MeCN_rot-isos_1996} contained besides their new 
measurements of the rotational spectra for the two singly substituted 
$^{13}$C isotopomers data from \citet{MeCN-isos_1979} for both species, 
covering 71$-$184~GHz as well as data for the $J = 2 - 1$ transitions 
for CH$_3^{13}$CN from \citet{MeCN-12-13b_2-1}.

The terrestrial $^{12}$C/$^{13}$C ratio is about 90. Hence, transitions 
of these species were easily observable in samples of natural isotopic 
composition, see Fig.~\ref{isos}. It is worthwhile mentioning that transitions 
of the main isotopolog in excited vibrational states 
$\varv _8 = 1$ and 2 and $\varv _4 = 1$ 
having the same $K$ are about as strong or even stronger 
at room temperature than the transitions of the $^{13}$C isotopomers. 

262 and 244 transitions were used in the final fits of $^{13}$CH$_3$CN 
and CH$_3^{13}$CN from the broad spectral recordings. $J''$ was extended to 
66 and 64, respectively, near 1.2~THz and $K$ reached 16 and 15, respectively. 
Some transitions were even identified near 1.6~THz, but they were too few 
and of too poor S/N to have a signifant impact on the data sets.
The weighted standard deviation for the lines measured at JPL are 
0.854 and 0.875, respectively, indicating essentially appropriate 
error estimates on average.

Only five transions each were recorded in Cologne for $^{13}$CH$_3$CN 
and CH$_3^{13}$CN at $J'' = 18$ and 16, respectively, and $K = 0 - 3$ 
and 6 in order to improve some of the low order spectroscopic parameters. 
Only in the process of preparing the manuscript was it noted that the 
residuals between measured frequencies and those calculated from the 
final set of spectroscopic parameters were considerably smaller than 
the assumed uncertainties of 3~kHz. In fact, when these transitions 
were weighted out the average residuals were only 1.9 and 1.5~kHz, 
respectively. 
Therefore, uncertainties of 1~kHz were assigned to these transitions. 
The resulting weighted standard deviations were then 0.303 and 0.571, 
respectively. This may indicate still too conservative error estimates, 
but because of the small number of lines which have much smaller 
uncertainties than most of the transition frequencies such small 
weighted standard deviations may not mean much.

The present and previous data were fit together to determine spectroscopic 
parameters for both isotopomers. Higher order parameters which could 
not be determined reliably were fixed to values derived from the main isotopic 
species by scaling the parameters with appropriate powers of $B$ and 
taking into consideration deviations from this estimate for related lower 
order parameters. The results are also given in Table~\ref{parameter_C3v}. 
The newly measured transition frequencies are given in 
Tables~\ref{tab:13CH3CN} and \ref{tab:CH3C-13-N} toward the end 
of the manuscript together with their assignments, uncertainties, 
and the residuals between measured frequencies and those 
calculated from the final spectroscopic parameters.

The rotational and centrifugal distortion parameters have been improved 
considerably with respect to \citet{MeCN_rot-isos_1996} while the values 
for $eQq$ are determined entirely from data used in that work. 
The weighted standard deviations for the whole fits are slightly below 1.0.


\subsection{CH$_3$C$^{15}$N}
\label{15N}

\citet{MeCN_rot-isos_1996} had in their data set transitions from 
\citet{MeCN-15N} covering 35$-$143~GHz and from \citet{MeCN-isos_1979} 
covering 214$-$232~GHz besides their own measurements. In the present fits 
we have used the $J = 1 - 0$ transition frequency from \citet{MeCN-15_1-0}.

The terrestrial $^{14}$N/$^{15}$N ratio is about 270, just a factor of three 
larger than the terrestrial $^{12}$C/$^{13}$C ratio. Thus, transitions of 
this isotopolog were still quite easily observable in samples of natural 
isotopic composition, see Fig.~\ref{isos}.
Eight transitions, $K = 0 - 6$ and 12 for $J'' = 18$ were recorded in Cologne. 
Additionally, 210 transition frequencies were extracted from the spectra 
recorded at JPL which extend to $J'' = 66$ near 1.2~THz and to $K = 14$. 

The present and previous data were fit together to determine spectroscopic 
parameters. As in subsection~\ref{13C}, higher order parameters were 
estimated and kept fixed. The spectroscopic parameters are given in 
Table~\ref{parameter_C3v}, too. The newly measured transition frequencies 
are given in Table~\ref{tab:CH3C15N}, again toward the end of the 
manuscript, together with their assignments, uncertainties, 
and the residuals between measured frequencies and those 
calculated from the final spectroscopic parameters.

The overall weighted standard deviation is 0.692, and the corresponding 
values for the transitions recorded in Cologne and at JPL are 0.833 and 
0.716.


\subsection{CH$_2$DCN}
\label{D}

\citet{CH2DCN-rot} presented extensive measurements of the singly deuterated 
methyl cyanide molecule between 116 and 471~GHz. They were able to observe 
several of the very weak $b$-type transitions because they employed an 
isotopically enriched sample. Their data set also included three 
$J = 2 - 1$ transitions from \citet{CH2DCN_2-1}. 

With transitions of CH$_3$C$^{15}$N identified almost as frequently 
as the singly substituted $^{13}$C species it seemed promising to 
detect CH$_2$DCN in samples of natural isotopic composition. 
The terrestrial H/D ratio is about 6400. The presence of three equivalent 
H atoms decreases the H/D ratio by three, the resolved $K$-doubling 
for low $K \ge 1$ increases it again by two, and the omission of 
spin-statistics complicates the situation even more. Altogether, the 
lines of CH$_2$DCN are around one order of magnitude weaker than the 
lines of CH$_3$C$^{15}$N. 

10 lines of 16 transitions with $J'' = 15 - 18$ and $K \le 10$ were recorded 
in Cologne. 109 lines of 152 transitions with $J'' \le 68$ and $K \le 11$ 
were extracted from the spectra taken at JPL. With increasing values of $J$ 
in the fit it was increasingly difficult to fit the data within experimental 
uncertainties. It turned out that  omission of one of the two very weak 
$K = 2 \leftrightarrow 1$ transitions reported by \citet{CH2DCN-rot} 
yielded satisfactory fits. Moreover, if one of these transitions 
was weighted out it appeared as if a 1~MHz typographical error may 
have occured for the other transition. Since the authors of this work 
did not have the spectra at their disposal and thus were unable to clarify 
this issue we decided to omit both $K = 2 \leftrightarrow 1$ transitions 
from the fit. Without these transitions, $D_K$ could not be determined 
reliably. Therefore, its value has been estimated by scaling values from 
ab initio calculations by the ratios of experimental versus ab initio 
values for CH$_3$CN (Table~\ref{parameter_C3v}) and CHD$_2$CN \citep{CHD2CN-rot}.
The value of 1.97~MHz is slightly bigger than the 1.83~MHz obtained 
in \citet{CH2DCN-rot}. $H_K$ was estimated as $D_K^2/A$ as done for 
CH$_3$CN, see subsection~\ref{main-iso}. The resulting spectroscopic 
parameters are given in Table~\ref{parameter_1D}.
The newly measured transition frequencies 
are given in Table~\ref{tab:CH2DCN}, toward the end of the 
manuscript, together with their assignments, uncertainties, 
and the residuals between measured frequencies and those 
calculated from the final spectroscopic parameters.

The weighted standard deviation for the whole fit is 0.740, for the 
Cologne and JPL lines it is 0.533 and 0.792, respectively. In particular 
the uncertainties of the Cologne lines have been estimated somewhat 
too conservatively.


\begin{table}
\begin{center}
\caption{Spectroscopic parameters$^a$ (MHz) of monodeuterated methyl 
cyanide, CH$_2$DCN and dimensionless weighted standard deviation wrms.}
\label{parameter_1D}
\renewcommand{\arraystretch}{1.10}
\begin{tabular}[t]{lr@{}l}
\hline \hline
Parameter & \multicolumn{2}{c}{Value} \\
\hline
$A$                      & 121\,074&.677\,(20)       \\
$B$                      &   8\,759&.197\,94\,(23)   \\
$C$                      &   8\,608&.542\,55\,(24)   \\
$D_K$                    &        1&.97              \\
$D_{JK} \times 10^3$     &      143&.336\,1\,(138)   \\
$D_J \times 10^3$        &        3&.469\,982\,(137) \\
$d_1 \times 10^6$        &    $-$79&.164\,(46)       \\
$d_2 \times 10^6$        &     $-$4&.776\,(35)       \\
$H_K \times 10^6$        &      100&.                \\
$H_{KJ} \times 10^6$     &        1&.688\,(192)      \\
$H_{JK} \times 10^9$     &      837&.5\,(49)         \\
$H_J \times 10^{12}$     &      400&.4\,(191)        \\
$h_1 \times 10^{12}$     &      149&.0\,(84)         \\
$L_{KKJ} \times 10^9$    &    $-$10&.82\,(68)        \\
$L_{JK} \times 10^{12}$  &       58&.\,(28)          \\
$L_{JJK} \times 10^{12}$ &     $-$5&.37\,(55)        \\
wrms                     &        0&.740             \\
\hline
\end{tabular}\\[2pt]
\end{center}
$^a$ Watson's $S$ reduction has been used in the representation $I^r$. 
Numbers in parentheses are one standard deviation in units of the least 
significant figures. Parameters without quoted uncertainties have been 
estimated; see section~\ref{a_and_d}.\\
\end{table}


\begin{figure}
\includegraphics[angle=0,width=9cm]{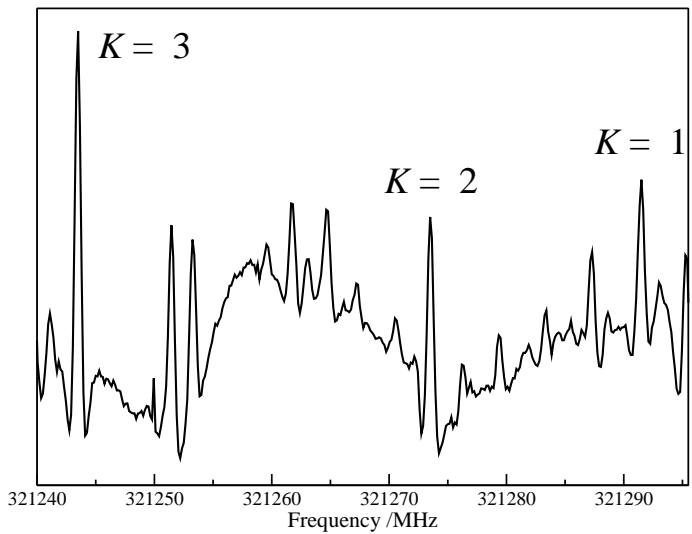}
  \caption{Section of the lower submillimeter spectrum of CH$_3$CN. 
   Transitions of $^{13}$CH$_3^{13}$CN have been labeled with their 
   $K$ quantum numbers. This isotopolog is less abundant by a factor 
   of 8000 compared to the main isotopolog. 
   While the spectrum is rather sparse on the level of the strongest 
   CH$_3$CN transitions, it is considerably less sparse on the level 
   of the CH$_3$C$^{15}$N transition, see Fig.~\ref{isos}, and rather 
   dense on the level of the $^{13}$CH$_3^{13}$CN transitions. 
   One of the lines may be due to a high-$K$, $\varv _8 = 1$ line of 
   $^{13}$CH$_3$CN, all other transitions remain unassigned, see 
   subsection~\ref{2x13}.}
\label{2x13-fig}
\end{figure}



\subsection{$^{13}$CH$_3^{13}$CN}
\label{2x13}

With even a considerable number of transitions assigned for CH$_2$DCN 
it seemed promising to assign transitions of the doubly substituted 
$^{13}$C isotopolog, as the lines are weaker by only about a factor of 
two. This species should be observable fairly readily in Galactic center 
sources because of the $^{12}$C/$^{13}$C ratio of about 20 in these sources 
\citep{isotopic_abundances,13C-VyCN_2008}. Hence, lines of the doubly 
substituted $^{13}$C isotopolog should be only slightly weaker than 
those of CH$_3$C$^{15}$N. In fact, inspection of the 3~mm line survey 
of Sagittarius~B2(N) obtained with the IRAM 30~m telecope (A.~Belloche, 
private communication, 2008; see also \citep{det-PrCN_EtFo,13C-VyCN_2008}) 
revealed that several $^{13}$CH$_3^{13}$CN lines should have S/N ratios 
of greater than three. However, all lines were overlapped by stronger lines, 
frequently  belonging to the much stronger $^{13}$CH$_3$CN which has 
fairly similar spectroscopic parameters. At higher frequencies, this 
overlap will be less severe.

\citet{MeCN-2x13} had reported some transition frequencies for this 
isotopolog up to 72~GHz. Inspection of the broad spectral recordings 
however failed to produce any even tentative assignments. Subsequently, 
all spectroscopic parameters were estimated from the isotopologs with 
no or only one $^{13}$C, and only $B$ was determined. Still, no 
assignments could be made.

The search for transitions of $^{13}$CH$_3^{13}$CN was the main reason 
for the measurements in Cologne. Several tentative assignments could be 
made near 321~GHz. All of these transitions were about 2.8~MHz higher 
than predicted, suggesting the assumptions of the centrifugal distortion 
parameters to be good, but the $B$ value to be slightly too small. 

With an abundance almost four orders of magnitude smaller than that of 
the main isotopolog, there are ample chances for overlap. 
Not only are there many vibrational states of the main isotopolog up to 
2000~cm$^{-1}$ which are about as strong or even stronger, but also 
transitions up to $\varv _4 = 1$ of the $^{13}$C isotopomers, or up 
to $\varv _8 = 1$ of the $^{15}$N species are about as strong. 
In addition, many additional species are only slightly weaker, 
including those of the isotopomers with $^{15}$N and one $^{13}$C. 
Even though the rotational spectrum of methyl cyanide is very sparse 
at the level of the strongest lines of the main isotopolog it is 
very rich at the level of the $^{13}$CH$_3^{13}$CN lines, as can be seen 
in Fig~\ref{2x13-fig}. Only the marked lines belonging to this isotopolog 
can be assigned unambiguously in this recording. One of the weaker lines 
may be due to a high-$K$, $\varv _8 = 1$ line of $^{13}$CH$_3$CN, 
but because of perturbations by $\varv _8 = 2$, the exact position of 
this transition is not certain. All other lines can not be assigned at present.

Altogether, 21 transitions with $J'' = 13 - 18$ and $K \le 6$ were 
recorded in Cologne. Subsequently, 34 transition with $J'' = 25 - 63$ 
and $K \le 9$ could be identified in the spectra taken at JPL. The data 
from \citet{MeCN-2x13} were omitted from the final fits because their 
large uncertainties cause negligible effects on the spectroscopic 
parameters.

Five parameters up to sixth order could be determined with significance; 
several other were kept fixed to estimated values as done further above 
(see subsection~\ref{13C}). Again, the spectroscopic parameters are 
listed in Table~\ref{parameter_C3v}.
The measured transition frequencies are given in Table~\ref{tab:2x13}, 
toward the end of the manuscript, together with their assignments, 
uncertainties, and the residuals between measured frequencies and 
those calculated from the final spectroscopic parameters.

The weighted standard deviation of the fit is 0.635. It is 0.719 and 
0.577 for the data from Cologne and JPL, respectively. The somewhat 
conservative error estimates can be justified by the possibility that 
some of the lines may still be affected by overlap to a non-negligible 
amount. However, the moderately large size of the data set should 
ensure that such effects are small. 

\section{Conclusion}
\label{conclusion}

Rotational transitions for six astrophysically and astrochemically 
important isotopologs of methyl cyanide in their ground vibrational 
states have been recorded and the existing data sets have been extended 
greatly in most cases. The data should permit reliable predictions 
to above 2~THz in all instances, sufficient not only for Herschel or 
ALMA, but very likely also for such missions as SOFIA, CCAT, or others.

Transitions of the even rarer isotopomers containing $^{15}$N and one 
$^{13}$C may be observable also, but chance of significant overlap 
are even higher and the chances to observe these species in space 
seem vanishingly low.

Predictions of the rotational spectra of the isotopologs studied in the 
course of the present investigation will be available in the catalog 
section\footnote{website: https://cdms.astro.uni-koeln.de/classic/entries/, 
see also https://cdms.astro.uni-koeln.de/classic/} 
of the Cologne Database for Molecular 
Spectroscopy\footnote{website: https://cdms.astro.uni-koeln.de/}
\citep{CDMS_1,CDMS_2}. The complete line, parameter and fit files 
will be deposited in the Spectroscopy Data section of the CDMS. 
Updated or new JPL catalog \citep{JPL-catalog} entries\footnote{website:  
http://spec.jpl.nasa.gov/ftp/pub/catalog/catdir.html, see also 
http://spec.jpl.nasa.gov/} will be available also.



\begin{acknowledgements}
H.S.P.M. is very grateful to the Bundesministerium f\"ur Bildung und 
Forschung (BMBF) for financial support aimed at maintaining the 
Cologne Database for Molecular Spectroscopy, CDMS. This support has been 
administered by the Deutsches Zentrum f\"ur Luft- und Raumfahrt (DLR). 
A part of the present research has been carried out at the Jet Propulsion 
Laboratory, California Institute of Technology, under a contract with the 
National Aeronautics and Space Administration (NASA).
\end{acknowledgements}


\clearpage
\onecolumn


\begin{center}
\small{
\begin{longtable}{rrr@{}lrrlrrr@{}lrrlrrr@{}lrr}
\caption{\label{tab:13CH3CN} Lower state quantum numbers of rotational transitions 
of $^{13}$CH$_3$CN from present work, frequencies (MHz), uncertainties unc. (kHz), and 
residuals o$-$c (kHz) between observed frequencies and those calculated from 
the final set of spectroscopic parameters.} \\
\hline \hline
$J''$ & $K''$ & \multicolumn{2}{c}{frequency} & unc. & o$-$c & &
$J''$ & $K''$ & \multicolumn{2}{c}{frequency} & unc. & o$-$c & &
$J''$ & $K''$ & \multicolumn{2}{c}{frequency} & unc. & o$-$c \\
\cline{1-6} \cline{8-13} \cline{15-20}
\endfirsthead
\caption{continued.} \\
\hline \hline
$J''$ & $K''$ & \multicolumn{2}{c}{frequency} & unc. & o$-$c & &
$J''$ & $K''$ & \multicolumn{2}{c}{frequency} & unc. & o$-$c & &
$J''$ & $K''$ & \multicolumn{2}{c}{frequency} & unc. & o$-$c \\
\cline{1-6} \cline{8-13} \cline{15-20}
\endhead
\hline
\endfoot
 18 &  6 &  339137.&3433  & 1.0 &  $-$0.3  & & 43 &  0 &  784895.&847   &  50 &      43  & & 56 &  3 & 1015544.&407   &  20 &      21  \\
 18 &  3 &  339309.&0082  & 1.0 &     0.3  & & 44 & 12 &  800553.&654   & 150 &  $-$143  & & 56 &  2 & 1015636.&946   &  30 &      17  \\
 18 &  2 &  339340.&8330  & 1.0 &     0.4  & & 44 & 11 &  800891.&319   & 100 &  $-$105  & & 56 &  1 & 1015692.&486   &  30 &      17  \\
 18 &  1 &  339359.&9326  & 1.0 &  $-$0.0  & & 44 & 10 &  801200.&157   &  50 &      13  & & 56 &  0 & 1015711.&009   &  50 &      24  \\
 18 &  0 &  339366.&2998  & 1.0 &  $-$0.4  & & 44 &  9 &  801479.&794   &  50 &   $-$44  & & 57 & 12 & 1030736.&928   & 150 &      51  \\
 24 & 12 &  445241.&735   &  30 &   $-$11  & & 44 &  8 &  801730.&374   &  50 &   $-$18  & & 57 & 10 & 1031558.&219   &  50 &      63  \\
 24 & 11 &  445432.&247   &  50 &      73  & & 44 &  7 &  801951.&637   &  50 &   $-$72  & & 57 &  8 & 1032231.&921   &  30 &      22  \\
 24 & 10 &  445606.&353   &  50 &      51  & & 44 &  6 &  802143.&651   &  50 &   $-$47  & & 57 &  7 & 1032513.&150   &  20 &      46  \\
 24 &  9 &  445764.&100   &  50 &      41  & & 44 &  5 &  802306.&261   &  50 &   $-$22  & & 57 &  6 & 1032757.&054   &  20 &      10  \\
 24 &  8 &  445905.&405   &  30 &      22  & & 44 &  4 &  802439.&367   &  50 &   $-$30  & & 57 &  5 & 1032963.&632   &  20 &      11  \\
 24 &  7 &  446030.&264   &  30 &      48  & & 44 &  3 &  802542.&979   &  50 &    $-$7  & & 57 &  4 & 1033132.&760   &  20 &       7  \\
 24 &  6 &  446138.&527   &  20 &      19  & & 44 &  2 &  802616.&938   &  50 &   $-$71  & & 57 &  2 & 1033358.&435   &  20 &      14  \\
 24 &  5 &  446230.&222   &  20 &       8  & & 44 &  1 &  802661.&445   &  50 &      10  & & 57 &  1 & 1033414.&879   &  20 &      12  \\
 24 &  4 &  446305.&355   &  50 &      56  & & 44 &  0 &  802676.&201   &  50 &   $-$44  & & 57 &  0 & 1033433.&720   &  20 &      35  \\
 24 &  3 &  446363.&747   &  20 &      17  & & 45 &  9 &  819230.&822   & 100 &  $-$149  & & 60 & 12 & 1083745.&465   &  50 &   $-$34  \\
 24 &  2 &  446405.&494   &  20 &      11  & & 45 &  8 &  819486.&798   & 100 &   $-$45  & & 60 & 11 & 1084195.&000   &  50 &    $-$0  \\
 24 &  1 &  446430.&566   &  20 &      24  & & 45 &  7 &  819712.&787   &  50 &   $-$70  & & 60 &  9 & 1084978.&318   &  50 &   $-$54  \\
 24 &  0 &  446438.&899   &  20 &       3  & & 45 &  6 &  819908.&940   &  50 &      20  & & 60 &  8 & 1085311.&933   &  50 &    $-$4  \\
 25 & 12 &  463032.&948   &  50 &      70  & & 45 &  5 &  820075.&010   &  50 &      54  & & 60 &  6 & 1085862.&162   &  50 &    $-$1  \\
 25 & 11 &  463230.&857   &  50 &      42  & & 45 &  4 &  820210.&775   &  50 &  $-$120  & & 60 &  4 & 1086255.&771   &  50 &   $-$44  \\
 25 & 10 &  463411.&809   &  80 &       2  & & 45 &  3 &  820316.&536   &  50 &  $-$147  & & 60 &  3 & 1086393.&678   &  50 &   $-$40  \\
 25 &  9 &  463575.&821   &  50 &      38  & & 45 &  2 &  820392.&237   &  50 &   $-$39  & & 60 &  2 & 1086492.&216   &  50 &   $-$45  \\
 25 &  6 &  463965.&038   &  50 &      45  & & 45 &  1 &  820437.&489   &  50 &  $-$156  & & 61 & 13 & 1100909.&779   & 100 &      92  \\
 25 &  5 &  464060.&345   &  50 &      31  & & 45 &  0 &  820452.&748   &  50 &   $-$22  & & 61 & 12 & 1101404.&872   &  50 &      11  \\
 25 &  4 &  464138.&400   &  50 &      41  & & 46 &  9 &  836978.&116   &  50 &   $-$59  & & 61 & 10 & 1102278.&334   &  80 &   $-$15  \\
 25 &  3 &  464199.&111   &  50 &      18  & & 46 &  8 &  837239.&242   &  80 &  $-$108  & & 61 &  9 & 1102656.&306   &  30 &   $-$19  \\
 25 &  2 &  464242.&534   &  50 &      42  & & 46 &  7 &  837469.&939   &  50 &  $-$107  & & 61 &  7 & 1103293.&992   &  20 &    $-$5  \\
 25 &  1 &  464268.&543   &  50 &       3  & & 46 &  6 &  837670.&126   &  50 &   $-$46  & & 61 &  6 & 1103553.&424   &  20 &   $-$17  \\
 25 &  0 &  464277.&228   &  50 &       5  & & 46 &  5 &  837839.&510   &  50 &  $-$137  & & 61 &  5 & 1103773.&128   &  50 &   $-$18  \\
 26 & 13 &  480598.&910   & 200 &      15  & & 46 &  4 &  837978.&301   &  50 &  $-$102  & & 61 &  4 & 1103953.&032   &  20 &       5  \\
 26 & 12 &  480821.&850   & 100 &      20  & & 46 &  3 &  838086.&283   &  50 &   $-$99  & & 61 &  3 & 1104093.&012   &  20 &       4  \\
 26 & 10 &  481215.&078   &  50 &   $-$26  & & 46 &  2 &  838163.&526   &  50 &   $-$15  & & 61 &  2 & 1104193.&040   &  20 &       4  \\
 26 &  9 &  481385.&284   &  50 &    $-$5  & & 46 &  1 &  838209.&741   &  50 &  $-$109  & & 61 &  1 & 1104253.&061   &  20 &    $-$8  \\
 26 &  8 &  481537.&813   &  50 &      67  & & 46 &  0 &  838225.&208   &  50 &   $-$81  & & 61 &  0 & 1104273.&082   &  20 &    $-$0  \\
 26 &  7 &  481672.&435   &  50 &      22  & & 49 & 12 &  889171.&685   &  80 &      46  & & 62 & 13 & 1118556.&496   & 100 &       4  \\
 26 &  6 &  481789.&313   &  50 &      78  & & 49 & 11 &  889544.&894   & 100 &      15  & & 62 & 12 & 1119059.&047   &  50 &      54  \\
 26 &  5 &  481888.&243   & 100 &      77  & & 49 & 10 &  889886.&195   &  80 &      32  & & 62 & 11 & 1119522.&099   &  80 &      76  \\
 26 &  4 &  481969.&220   &  50 &      55  & & 49 &  9 &  890195.&361   &  80 &       5  & & 62 & 10 & 1119945.&421   &  40 &      16  \\
 26 &  3 &  482032.&241   &  50 &      42  & & 49 &  8 &  890472.&317   &  50 &   $-$20  & & 62 &  8 & 1120672.&581   &  20 &       6  \\
 26 &  2 &  482077.&319   &  50 &      78  & & 49 &  7 &  890716.&982   &  50 &   $-$12  & & 62 &  7 & 1120976.&075   &  20 &    $-$1  \\
 26 &  1 &  482104.&227   &  50 &   $-$48  & & 49 &  6 &  890929.&260   &  50 &      28  & & 62 &  6 & 1121239.&357   &  50 &    $-$1  \\
 26 &  0 &  482113.&311   &  50 &      24  & & 49 &  5 &  891108.&967   &  50 &       4  & & 62 &  5 & 1121462.&318   &  20 &       5  \\
 27 & 12 &  498608.&539   & 150 &      24  & & 49 &  4 &  891256.&131   &  50 &      16  & & 62 &  3 & 1121786.&918   &  20 &      12  \\
 27 & 11 &  498821.&247   & 150 &  $-$178  & & 49 &  3 &  891370.&650   &  50 &      21  & & 62 &  2 & 1121888.&415   &  20 &       1  \\
 27 & 10 &  499016.&263   & 100 &     151  & & 49 &  2 &  891452.&462   &  50 &       5  & & 62 &  1 & 1121949.&341   &  20 &       7  \\
 27 &  8 &  499350.&520   &  50 &      18  & & 49 &  1 &  891501.&579   &  50 &      12  & & 62 &  0 & 1121969.&656   &  30 &      12  \\
 27 &  7 &  499490.&120   &  50 &      47  & & 49 &  0 &  891517.&973   &  50 &      33  & & 63 & 15 & 1135059.&113   & 150 &      40  \\
 27 &  6 &  499611.&153   &  50 &       4  & & 50 & 13 &  906470.&254   & 150 &      42  & & 63 & 11 & 1137177.&559   &  50 &    $-$3  \\
 27 &  5 &  499713.&686   &  50 &       4  & & 50 & 12 &  906882.&928   &  50 &       9  & & 63 & 10 & 1137607.&095   &  30 &       4  \\
 27 &  4 &  499797.&648   &  50 &      17  & & 50 & 11 &  907263.&229   &  80 &      15  & & 63 &  9 & 1137996.&224   &  20 &    $-$3  \\
 27 &  2 &  499909.&654   &  50 &      11  & & 50 & 10 &  907610.&895   & 100 &   $-$53  & & 63 &  8 & 1138344.&827   &  50 &       9  \\
 27 &  1 &  499937.&687   &  50 &      27  & & 50 &  9 &  907925.&987   &  50 &       1  & & 63 &  7 & 1138652.&740   &  20 &      15  \\
 27 &  0 &  499947.&025   &  50 &      24  & & 50 &  8 &  908208.&205   &  50 &       3  & & 63 &  6 & 1138919.&837   &  20 &       8  \\
 28 & 11 &  516613.&238   &  80 &      10  & & 50 &  7 &  908457.&487   &  80 &       4  & & 63 &  5 & 1139146.&023   &  20 &       2  \\
 28 & 10 &  516814.&764   &  80 &      23  & & 50 &  6 &  908673.&729   &  50 &    $-$2  & & 63 &  4 & 1139331.&185   &  50 &   $-$27  \\
 28 &  9 &  516997.&294   &  50 &   $-$16  & & 50 &  5 &  908856.&837   &  50 &   $-$22  & & 63 &  3 & 1139475.&324   &  20 &    $-$2  \\
 28 &  3 &  517691.&313   &  50 &      22  & & 50 &  4 &  909006.&769   &  50 &   $-$23  & & 63 &  2 & 1139578.&300   &  20 &    $-$7  \\
 28 &  2 &  517739.&642   &  50 &      31  & & 50 &  3 &  909123.&446   &  50 &   $-$23  & & 63 &  1 & 1139640.&092   &  20 &   $-$19  \\
 28 &  1 &  517768.&629   &  50 &      19  & & 50 &  2 &  909206.&838   &  50 &    $-$7  & & 64 & 14 & 1153276.&801   & 200 &     131  \\
 28 &  0 &  517778.&310   &  50 &      31  & & 50 &  1 &  909256.&926   &  50 &      43  & & 64 & 12 & 1154351.&199   &  30 &   $-$21  \\
 34 & 16 &  621760.&772   &  50 &      10  & & 50 &  0 &  909273.&539   &  50 &   $-$26  & & 64 &  9 & 1155657.&985   &  20 &   $-$18  \\
 34 & 14 &  622447.&422   &  30 &   $-$26  & & 51 & 12 &  924589.&926   &  80 &      22  & & 64 &  8 & 1156011.&566   &  20 &       4  \\
 34 & 11 &  623309.&671   &  20 &      14  & & 51 & 11 &  924977.&218   & 100 &   $-$13  & & 64 &  7 & 1156323.&830   &  20 &   $-$27  \\
 34 & 10 &  623551.&857   &  20 &       5  & & 51 &  9 &  925652.&265   &  50 &       8  & & 64 &  6 & 1156594.&758   &  20 &    $-$9  \\
 34 &  9 &  623771.&262   &  20 &   $-$14  & & 51 &  8 &  925939.&684   &  50 &    $-$7  & & 64 &  4 & 1157012.&016   &  20 &       5  \\
 34 &  7 &  624141.&443   &  20 &   $-$28  & & 51 &  7 &  926193.&576   &  80 &    $-$5  & & 64 &  3 & 1157158.&166   &  20 &   $-$13  \\
 34 &  6 &  624292.&087   &  20 &    $-$5  & & 51 &  5 &  926600.&341   &  80 &       2  & & 64 &  2 & 1157262.&634   &  20 &       8  \\
 34 &  4 &  624524.&053   &  20 &   $-$26  & & 51 &  4 &  926753.&052   &  80 &       8  & & 64 &  0 & 1157346.&208   &  20 &    $-$2  \\
 34 &  2 &  624663.&446   &  20 &      24  & & 51 &  3 &  926871.&878   &  50 &    $-$0  & & 65 & 13 & 1171464.&851   & 100 &      10  \\
 34 &  1 &  624698.&263   &  20 &   $-$13  & & 51 &  2 &  926956.&781   &  50 &   $-$14  & & 65 & 11 & 1172472.&160   &  80 &  $-$102  \\
 34 &  0 &  624709.&879   &  20 &   $-$16  & & 51 &  1 &  927007.&756   &  50 &    $-$2  & & 65 &  9 & 1173314.&209   &  30 &    $-$5  \\
 35 & 16 &  639490.&666   & 100 &      54  & & 51 &  0 &  927024.&756   &  50 &       8  & & 65 &  8 & 1173672.&712   &  20 &    $-$8  \\
 35 & 14 &  640196.&404   &  60 &      27  & & 55 & 12 &  995373.&248   & 100 &      44  & & 65 &  6 & 1174264.&072   &  20 &   $-$12  \\
 35 & 12 &  640810.&267   &  30 &   $-$44  & & 55 & 10 &  996168.&138   & 100 &      60  & & 65 &  4 & 1174687.&142   &  50 &   $-$23  \\
 35 & 11 &  641082.&493   &  50 &   $-$47  & & 55 &  9 &  996511.&966   &  50 &   $-$72  & & 65 &  3 & 1174835.&374   &  20 &    $-$3  \\
 35 &  9 &  641556.&961   &  50 &   $-$22  & & 55 &  7 &  997092.&349   &  20 &      21  & & 65 &  2 & 1174941.&285   &  20 &    $-$1  \\
 35 &  7 &  641937.&464   &  20 &       1  & & 55 &  5 &  997528.&396   &  20 &      33  & & 65 &  1 & 1175004.&858   &  20 &       9  \\
 35 &  6 &  642092.&248   &  50 &   $-$20  & & 55 &  4 &  997692.&055   &  20 &    $-$4  & & 65 &  0 & 1175026.&016   &  30 &   $-$23  \\
 35 &  5 &  642223.&312   &  50 &   $-$53  & & 55 &  3 &  997819.&444   &  30 &    $-$3  & & 66 & 12 & 1189621.&606   & 120 &     111  \\
 35 &  4 &  642330.&670   &  30 &   $-$28  & & 55 &  2 &  997910.&473   &  20 &    $-$2  & & 66 & 11 & 1190111.&236   & 150 &   $-$17  \\
 35 &  3 &  642414.&174   &  50 &   $-$52  & & 55 &  1 &  997965.&147   &  30 &      40  & & 66 & 10 & 1190558.&963   & 120 &  $-$107  \\
 35 &  2 &  642473.&884   &  30 &   $-$29  & & 55 &  0 &  997983.&325   &  20 &       6  & & 66 &  9 & 1190964.&780   &  20 &       6  \\
 35 &  0 &  642521.&653   &  30 &   $-$24  & & 56 & 12 & 1013057.&279   & 100 &  $-$164  & & 66 &  6 & 1191927.&684   &  25 &   $-$10  \\
 43 &  8 &  783970.&101   &  80 &      18  & & 56 &  9 & 1014215.&234   &  20 &       7  & & 66 &  5 & 1192163.&522   &  20 &       8  \\
 43 &  6 &  784374.&602   &  50 &      11  & & 56 &  8 & 1014528.&429   &  50 &   $-$48  & & 66 &  4 & 1192356.&610   &  20 &      23  \\
 43 &  5 &  784533.&705   &  50 &   $-$10  & & 56 &  7 & 1014805.&211   &  20 &      40  & & 66 &  3 & 1192506.&829   &  20 &    $-$6  \\
 43 &  3 &  784765.&344   &  50 &   $-$36  & & 56 &  6 & 1015045.&237   &  30 &      40  & & 66 &  2 & 1192614.&182   &  20 &   $-$17  \\
 43 &  2 &  784837.&839   &  50 &      11  & & 56 &  5 & 1015248.&484   &  20 &      23  & & 66 &  1 & 1192678.&616   &  20 &   $-$19  \\
 43 &  1 &  784881.&204   &  50 &  $-$104  & & 56 &  4 & 1015414.&879   &  20 &    $-$1  & & 66 &  0 & 1192700.&114   &  20 &    $-$2  \\
\end{longtable}
}
\end{center}


\begin{center}
\small{
\begin{longtable}{rrr@{}lrrlrrr@{}lrrlrrr@{}lrr}
\caption{\label{tab:CH3C-13-N} Lower state quantum numbers of rotational transitions 
of CH$_3^{13}$CN from present work, frequencies (MHz), uncertainties unc. (kHz), and 
residuals o$-$c (kHz) between observed frequencies and those calculated from 
the final set of spectroscopic parameters.} \\
\hline \hline
$J''$ & $K''$ & \multicolumn{2}{c}{frequency} & unc. & o$-$c & &
$J''$ & $K''$ & \multicolumn{2}{c}{frequency} & unc. & o$-$c & &
$J''$ & $K''$ & \multicolumn{2}{c}{frequency} & unc. & o$-$c \\
\cline{1-6} \cline{8-13} \cline{15-20}
\endfirsthead
\caption{continued.} \\
\hline \hline
$J''$ & $K''$ & \multicolumn{2}{c}{frequency} & unc. & o$-$c & &
$J''$ & $K''$ & \multicolumn{2}{c}{frequency} & unc. & o$-$c & &
$J''$ & $K''$ & \multicolumn{2}{c}{frequency} & unc. & o$-$c \\
\cline{1-6} \cline{8-13} \cline{15-20}
\endhead
\hline
\endfoot
 16 &  6 &  312317&.7600 & 1.0 &   0.1 & & 34 &  1 &  642938&.809  &  50 & $-$45 & & 53 &  3 &  990423&.344  &  50 & $-$29 \\
 16 &  3 &  312479&.1625 & 1.0 &   0.2 & & 34 &  0 &  642950&.994  &  50 & $-$54 & & 53 &  2 &  990515&.635  &  20 &  $-$1 \\
 16 &  2 &  312509&.0842 & 1.0 &   0.1 & & 41 &  6 &  770673&.135  &  50 & $-$47 & & 53 &  1 &  990571&.014  &  20 &     6 \\
 16 &  1 &  312527&.0409 & 1.0 &$-$1.1 & & 41 &  4 &  770963&.549  &  50 & $-$25 & & 53 &  0 &  990589&.492  &  20 &    24 \\
 16 &  0 &  312533&.0294 & 1.0 &   0.6 & & 41 &  3 &  771065&.296  &  50 &  $-$8 & & 54 &  6 & 1008167&.091  &  30 &    44 \\
 23 & 10 &  440278&.491  &  50 & $-$21 & & 41 &  2 &  771137&.916  &  50 & $-$81 & & 54 &  4 & 1008542&.013  &  30 &    25 \\
 23 &  9 &  440437&.626  &  20 &    20 & & 41 &  1 &  771181&.580  &  50 & $-$45 & & 54 &  2 & 1008767&.251  &  50 &    58 \\
 23 &  8 &  440580&.110  &  20 & $-$13 & & 41 &  0 &  771196&.159  &  50 & $-$11 & & 54 &  1 & 1008823&.503  &  30 & $-$19 \\
 23 &  7 &  440706&.004  &  20 &  $-$5 & & 42 & 10 &  788019&.367  & 150 &$-$162 & & 54 &  0 & 1008842&.357  &  50 &    57 \\
 23 &  6 &  440815&.253  &  20 &    40 & & 42 &  9 &  788300&.498  &  50 &    10 & & 55 &  9 & 1025547&.406  &  30 &    34 \\
 23 &  5 &  440907&.717  &  20 &    25 & & 42 &  8 &  788552&.113  &  50 & $-$61 & & 55 &  8 & 1025870&.459  &  50 &  $-$6 \\
 23 &  4 &  440983&.429  &  30 &    22 & & 42 &  7 &  788774&.377  &  50 &$-$108 & & 55 &  7 & 1026155&.838  &  50 & $-$13 \\
 23 &  1 &  441109&.693  &  20 &  $-$8 & & 42 &  6 &  788967&.211  &  50 &$-$125 & & 55 &  6 & 1026403&.441  &  50 &    25 \\
 23 &  0 &  441118&.148  &  20 &    23 & & 42 &  5 &  789130&.563  &  50 & $-$86 & & 55 &  4 & 1026784&.724  &  30 &    22 \\
 24 & 12 &  458222&.531  & 120 &    20 & & 42 &  4 &  789264&.223  &  80 &$-$135 & & 55 &  3 & 1026918&.341  &  50 &    68 \\
 24 & 11 &  458422&.402  & 120 & $-$48 & & 42 &  3 &  789368&.387  &  50 & $-$24 & & 55 &  2 & 1027013&.731  &  20 &    11 \\
 24 & 10 &  458605&.156  & 120 &$-$114 & & 42 &  2 &  789442&.628  &  80 &$-$135 & & 55 &  1 & 1027071&.016  &  20 &    14 \\
 24 &  7 &  459050&.448  &  50 &   118 & & 42 &  1 &  789487&.346  &  50 & $-$40 & & 55 &  0 & 1027090&.097  &  20 &  $-$2 \\
 24 &  6 &  459164&.066  &  50 &    45 & & 42 &  0 &  789502&.201  &  50 & $-$62 & & 56 & 12 & 1042550&.491  &  50 &    34 \\
 24 &  5 &  459260&.263  &  50 & $-$36 & & 43 & 12 &  805624&.829  &  80 &$-$124 & & 56 & 11 & 1042993&.029  &  50 & $-$55 \\
 24 &  4 &  459339&.106  &  50 & $-$18 & & 43 & 10 &  806288&.633  &  50 & $-$59 & & 56 & 10 & 1043397&.819  &  30 &    14 \\
 24 &  3 &  459400&.506  &  50 &    40 & & 43 &  9 &  806575&.857  &  50 & $-$47 & & 56 &  9 & 1043764&.461  &  30 &  $-$1 \\
 24 &  2 &  459444&.289  &  50 & $-$10 & & 43 &  8 &  806833&.106  &  50 & $-$84 & & 56 &  8 & 1044092&.961  &  20 &    46 \\
 24 &  1 &  459470&.550  &  50 & $-$57 & & 43 &  7 &  807060&.397  &  50 & $-$53 & & 56 &  7 & 1044383&.048  &  30 &    15 \\
 24 &  0 &  459479&.378  &  50 &     1 & & 43 &  6 &  807257&.542  &  50 & $-$50 & & 56 &  4 & 1045022&.335  &  30 &    24 \\
 25 & 12 &  476531&.946  &  80 & $-$16 & & 43 &  5 &  807424&.442  & 100 & $-$97 & & 56 &  2 & 1045255&.204  &  30 &    79 \\
 25 & 10 &  476929&.757  &  80 & $-$44 & & 43 &  4 &  807561&.186  &  50 & $-$37 & & 56 &  1 & 1045313&.390  &  20 &    32 \\
 25 &  9 &  477101&.989  &  50 &    35 & & 43 &  3 &  807667&.545  &  50 & $-$46 & & 56 &  0 & 1045332&.804  &  20 &    33 \\
 25 &  8 &  477256&.174  &  50 &     1 & & 43 &  2 &  807743&.522  &  50 & $-$77 & & 60 & 13 & 1114776&.860  &  50 & $-$32 \\
 25 &  7 &  477392&.405  &  50 &    11 & & 43 &  1 &  807789&.195  &  50 & $-$20 & & 60 & 12 & 1115288&.291  &  20 &  $-$4 \\
 25 &  6 &  477510&.574  &  50 &    11 & & 43 &  0 &  807804&.377  &  50 & $-$46 & & 60 & 10 & 1116190&.390  &  20 &  $-$1 \\
 25 &  5 &  477610&.680  &  50 &    46 & & 44 &  7 &  825342&.348  & 100 & $-$89 & & 60 &  9 & 1116580&.728  &  20 & $-$11 \\
 25 &  4 &  477692&.603  &  50 &    39 & & 44 &  6 &  825543&.773  &  50 & $-$85 & & 60 &  8 & 1116930&.416  &  20 &     6 \\
 25 &  3 &  477756&.356  &  50 &    33 & & 44 &  5 &  825714&.366  & 100 & $-$62 & & 60 &  7 & 1117239&.270  &  20 &  $-$0 \\
 25 &  2 &  477801&.869  &  50 & $-$14 & & 44 &  4 &  825853&.985  &  50 & $-$94 & & 60 &  6 & 1117507&.196  &  20 &  $-$2 \\
 25 &  1 &  477829&.257  &  50 &    30 & & 44 &  2 &  826040&.390  &  50 & $-$22 & & 60 &  5 & 1117734&.082  &  50 &  $-$4 \\
 25 &  0 &  477838&.379  &  50 &    37 & & 44 &  1 &  826086&.880  &  50 &$-$138 & & 60 &  4 & 1117919&.833  &  20 & $-$13 \\
 26 & 12 &  494839&.226  &  50 &   102 & & 45 & 12 &  842122&.594  & 150 & $-$45 & & 60 &  2 & 1118167&.698  &  20 &  $-$0 \\
 26 & 11 &  495054&.907  & 100 &   104 & & 45 & 10 &  842814&.997  & 120 &$-$150 & & 60 &  1 & 1118229&.675  &  20 & $-$17 \\
 26 & 10 &  495252&.014  &  50 &  $-$0 & & 45 &  9 &  843114&.715  &  80 & $-$93 & & 60 &  0 & 1118250&.348  &  20 & $-$12 \\
 26 &  9 &  495430&.686  &  50 &     4 & & 47 & 13 &  878195&.374  & 150 &    79 & & 61 &  9 & 1134771&.355  &  50 &    23 \\
 26 &  8 &  495590&.783  &  50 &    47 & & 47 & 12 &  878604&.078  &  80 &  $-$4 & & 61 &  8 & 1135126&.272  &  20 &    17 \\
 26 &  7 &  495732&.066  &  50 & $-$44 & & 47 & 11 &  878980&.835  &  80 &    76 & & 61 &  7 & 1135439&.732  &  30 & $-$20 \\
 26 &  6 &  495854&.766  &  50 &    16 & & 47 & 10 &  879325&.220  &  50 &    39 & & 61 &  6 & 1135711&.703  &  20 &  $-$1 \\
 26 &  4 &  496043&.684  &  50 &    47 & & 47 &  9 &  879637&.230  &  50 &    18 & & 61 &  5 & 1135941&.998  &  20 &  $-$1 \\
 26 &  3 &  496109&.798  &  50 &  $-$9 & & 47 &  8 &  879916&.758  &  50 &    27 & & 61 &  2 & 1136382&.127  &  20 &     6 \\
 26 &  2 &  496157&.109  &  50 &    18 & & 47 &  7 &  880163&.650  &  50 &    22 & & 61 &  1 & 1136445&.038  &  30 &  $-$8 \\
 26 &  1 &  496185&.478  &  50 &     8 & & 47 &  6 &  880377&.805  &  50 &     0 & & 61 &  0 & 1136466&.018  &  20 &  $-$6 \\
 26 &  0 &  496194&.930  &  50 &     0 & & 47 &  5 &  880559&.094  & 100 & $-$84 & & 62 & 15 & 1149921&.646  &  50 &     6 \\
 27 & 12 &  513143&.945  &  80 &    36 & & 47 &  4 &  880707&.708  &  50 &    35 & & 62 & 14 & 1150530&.560  &  50 &    37 \\
 27 & 10 &  513571&.914  & 100 &    91 & & 47 &  3 &  880823&.252  & 100 &    21 & & 62 & 12 & 1151625&.207  &  50 &    36 \\
 27 &  9 &  513757&.039  &  50 &    47 & & 47 &  2 &  880905&.815  &  50 &    10 & & 62 & 11 & 1152110&.498  &  30 & $-$25 \\
 27 &  8 &  513922&.895  &  50 &    26 & & 47 &  1 &  880955&.373  &  50 &    10 & & 62 & 10 & 1152554&.304  &  30 &  $-$3 \\
 27 &  7 &  514069&.366  &  50 & $-$21 & & 47 &  0 &  880971&.914  &  50 &    29 & & 62 &  8 & 1153316&.508  &  20 &     3 \\
 27 &  6 &  514196&.520  &  50 &    29 & & 48 & 13 &  896421&.629  & 120 &  $-$9 & & 62 &  7 & 1153634&.632  &  25 &    10 \\
 27 &  5 &  514304&.162  &  50 &    36 & & 48 & 12 &  896838&.549  &  80 &    59 & & 62 &  6 & 1153910&.575  &  20 &  $-$3 \\
 27 &  4 &  514392&.277  &  50 &    27 & & 48 & 10 &  897573&.823  &  80 &    10 & & 62 &  5 & 1154144&.275  &  20 &     9 \\
 27 &  3 &  514460&.821  &  50 &  $-$8 & & 48 &  9 &  897892&.002  &  50 &     2 & & 62 &  4 & 1154335&.590  &  20 &  $-$2 \\
 27 &  2 &  514509&.842  &  50 &     9 & & 48 &  8 &  898177&.045  &  50 &    13 & & 62 &  2 & 1154590&.861  &  20 & $-$11 \\
 27 &  1 &  514539&.243  &  50 &  $-$0 & & 48 &  7 &  898428&.827  &  50 &    27 & & 62 &  1 & 1154654&.718  &  20 &  $-$5 \\
 27 &  0 &  514549&.047  &  50 &  $-$1 & & 48 &  6 &  898647&.220  &  50 &    19 & & 62 &  0 & 1154676&.018  &  20 &     8 \\
 33 & 15 &  621969&.166  &  30 &    40 & & 48 &  4 &  898983&.584  &  50 &     9 & & 63 & 12 & 1169785&.281  &  20 &  $-$4 \\
 33 & 14 &  622308&.158  &  30 & $-$16 & & 48 &  2 &  899185&.623  &  50 &     8 & & 63 & 11 & 1170277&.655  &  30 &     0 \\
 33 & 13 &  622624&.402  &  20 & $-$16 & & 48 &  1 &  899236&.139  &  50 & $-$12 & & 63 & 10 & 1170727&.840  &  30 & $-$15 \\
 33 & 10 &  623435&.148  &  30 & $-$11 & & 48 &  0 &  899253&.011  &  50 &    13 & & 63 &  9 & 1171135&.718  &  40 &     5 \\
 33 &  9 &  623659&.045  &  20 & $-$13 & & 49 & 13 &  914643&.473  & 200 &$-$207 & & 63 &  7 & 1171823&.773  &  20 & $-$14 \\
 33 &  8 &  623859&.664  &  20 &    34 & & 49 & 10 &  915818&.023  & 150 & $-$47 & & 63 &  6 & 1172103&.723  &  20 &  $-$9 \\
 33 &  7 &  624036&.764  &  20 & $-$31 & & 49 &  9 &  916142&.366  &  50 & $-$25 & & 63 &  5 & 1172340&.795  &  25 &  $-$2 \\
 33 &  6 &  624190&.451  &  20 & $-$31 & & 49 &  8 &  916432&.928  &  50 &     8 & & 63 &  4 & 1172534&.868  &  40 & $-$21 \\
 33 &  5 &  624320&.650  &  20 &    21 & & 49 &  7 &  916689&.536  &  50 &  $-$4 & & 63 &  2 & 1172793&.840  &  25 & $-$19 \\
 33 &  4 &  624427&.221  &  30 &    36 & & 49 &  6 &  916912&.168  &  50 &    16 & & 63 &  1 & 1172858&.675  &  40 &    42 \\
 33 &  3 &  624510&.071  &  30 & $-$36 & & 49 &  4 &  917255&.016  &  50 &     5 & & 63 &  0 & 1172880&.216  &  20 & $-$11 \\
 33 &  2 &  624569&.356  &  20 &  $-$5 & & 49 &  2 &  917460&.920  &  50 & $-$26 & & 64 & 12 & 1187939&.602  & 120 &$-$128 \\
 33 &  0 &  624616&.735  &  30 & $-$42 & & 49 &  1 &  917512&.445  &  50 & $-$10 & & 64 & 11 & 1188438&.974  & 150 &$-$112 \\
 34 & 15 &  640227&.661  &  40 &     8 & & 49 &  0 &  917529&.610  &  50 & $-$17 & & 64 & 10 & 1188895&.674  &  30 &  $-$1 \\
 34 & 14 &  640576&.289  &  60 &$-$112 & & 53 & 13 &  987487&.235  & 200 &   170 & & 64 &  9 & 1189309&.325  &  20 &     4 \\
 34 & 13 &  640901&.691  &  30 &  $-$2 & & 53 & 11 &  988364&.813  &  80 &   104 & & 64 &  8 & 1189679&.851  &  25 & $-$11 \\
 34 & 11 &  641481&.367  &  50 & $-$52 & & 53 & 10 &  988749&.584  &  50 &    37 & & 64 &  7 & 1190007&.145  &  30 & $-$12 \\
 34 &  8 &  642172&.196  &  30 & $-$45 & & 53 &  9 &  989098&.196  &  20 &     3 & & 64 &  6 & 1190291&.073  &  20 &  $-$1 \\
 34 &  7 &  642354&.446  &  30 & $-$28 & & 53 &  8 &  989410&.512  &  20 &     1 & & 64 &  5 & 1190531&.497  &  20 &  $-$5 \\
 34 &  4 &  642756&.021  &  20 & $-$12 & & 53 &  7 &  989686&.393  &  20 &    15 & & 64 &  4 & 1190728&.334  &  30 & $-$13 \\
 34 &  3 &  642841&.300  &  20 & $-$26 & & 53 &  6 &  989925&.693  &  20 &     7 & & 64 &  2 & 1190990&.978  &  20 & $-$13 \\
 34 &  2 &  642902&.247  &  30 & $-$28 & & 53 &  4 &  990294&.279  &  20 &    22 & & 64 &  1 & 1191056&.621  &  50 & $-$63 \\
\end{longtable}
}
\end{center}


\begin{center}
\small{
\begin{longtable}{rrr@{}lrrlrrr@{}lrrlrrr@{}lrr}
\caption{\label{tab:CH3C15N} Lower state quantum numbers of rotational transitions 
of CH$_3$C$^{15}$N from present work, frequencies (MHz), uncertainties unc. (kHz), and 
residuals o$-$c (kHz) between observed frequencies and those calculated from 
the final set of spectroscopic parameters.} \\
\hline \hline
$J''$ & $K''$ & \multicolumn{2}{c}{frequency} & unc. & o$-$c & &
$J''$ & $K''$ & \multicolumn{2}{c}{frequency} & unc. & o$-$c & &
$J''$ & $K''$ & \multicolumn{2}{c}{frequency} & unc. & o$-$c \\
\cline{1-6} \cline{8-13} \cline{15-20}
\endfirsthead
\caption{continued.} \\
\hline \hline
$J''$ & $K''$ & \multicolumn{2}{c}{frequency} & unc. & o$-$c & &
$J''$ & $K''$ & \multicolumn{2}{c}{frequency} & unc. & o$-$c & &
$J''$ & $K''$ & \multicolumn{2}{c}{frequency} & unc. & o$-$c \\
\cline{1-6} \cline{8-13} \cline{15-20}
\endhead
\hline
\endfoot
 18 & 12 &  338023&.606 &  10 &     0 & & 35 & 11 &  640278&.287 &  70 & $-$75 & & 56 &  7 & 1013569&.196 &  30 &    51 \\
 18 &  6 &  338710&.019 &   8 &  $-$8 & & 35 &  9 &  640754&.725 &  30 &  $-$9 & & 56 &  6 & 1013809&.982 &  30 & $-$23 \\
 18 &  5 &  338780&.216 &   3 &     1 & & 35 &  7 &  641136&.691 &  50 & $-$38 & & 56 &  5 & 1014014&.014 &  50 &    45 \\
 18 &  4 &  338837&.680 &   3 &     1 & & 35 &  6 &  641292&.126 &  40 & $-$18 & & 56 &  4 & 1014180&.965 &  20 &     7 \\
 18 &  3 &  338882&.394 &   3 &  $-$4 & & 35 &  4 &  641531&.486 &  30 & $-$18 & & 56 &  3 & 1014310&.943 &  50 &    36 \\
 18 &  2 &  338914&.344 &   8 &  $-$8 & & 35 &  3 &  641615&.332 &  50 & $-$23 & & 56 &  2 & 1014403&.766 &  20 &     3 \\
 18 &  1 &  338933&.536 &   5 &     6 & & 35 &  2 &  641675&.251 &  20 & $-$22 & & 56 &  1 & 1014459&.492 &  30 &  $-$0 \\
 18 &  0 &  338939&.926 &   5 &     2 & & 35 &  1 &  641711&.248 &  20 &    16 & & 56 &  0 & 1014478&.090 &  50 &    19 \\
 24 & 12 &  444677&.654 &  50 &    68 & & 44 & 12 &  799556&.798 & 100 & $-$10 & & 57 &  8 & 1030975&.149 &  50 &    27 \\
 24 & 11 &  444868&.773 &  80 & $-$60 & & 44 & 11 &  799895&.712 &  80 & $-$58 & & 57 &  6 & 1031502&.077 &  20 &  $-$9 \\
 24 & 10 &  445043&.705 &  50 &     5 & & 44 &  9 &  800486&.471 &  50 &  $-$2 & & 57 &  5 & 1031709&.364 &  30 &  $-$6 \\
 24 &  9 &  445202&.122 &  50 &     2 & & 44 &  8 &  800737&.904 & 100 & $-$84 & & 57 &  4 & 1031879&.082 &  20 &     6 \\
 24 &  7 &  445469&.416 &  50 &    36 & & 44 &  7 &  800960&.070 &  50 & $-$74 & & 57 &  3 & 1032011&.155 &  20 &    15 \\
 24 &  6 &  445578&.097 &  50 & $-$19 & & 44 &  6 &  801152&.797 &  50 & $-$59 & & 57 &  2 & 1032105&.530 &  20 &    23 \\
 24 &  5 &  445670&.197 &  50 &     1 & & 44 &  5 &  801315&.982 &  50 & $-$66 & & 57 &  1 & 1032162&.147 &  20 &     4 \\
 24 &  4 &  445745&.587 &  50 &     4 & & 44 &  4 &  801449&.656 &  50 &  $-$1 & & 57 &  0 & 1032181&.021 &  20 &  $-$4 \\
 24 &  3 &  445804&.286 &  50 &    36 & & 44 &  3 &  801553&.593 &  50 & $-$37 & & 60 &  8 & 1083996&.412 &  30 & $-$21 \\
 24 &  2 &  445846&.202 &  50 &    31 & & 44 &  2 &  801627&.891 &  50 & $-$35 & & 60 &  7 & 1084292&.050 &  50 & $-$17 \\
 24 &  1 &  445871&.360 &  50 &    29 & & 44 &  1 &  801672&.512 &  50 &  $-$2 & & 60 &  5 & 1084765&.613 &  80 & $-$70 \\
 24 &  0 &  445879&.744 &  50 &    26 & & 44 &  0 &  801687&.390 &  50 &    10 & & 60 &  3 & 1085081&.878 &  50 &    41 \\
 25 & 11 &  462645&.335 & 100 &    10 & & 45 &  5 &  819063&.889 &  50 & $-$13 & & 60 &  1 & 1085240&.023 &  20 & $-$15 \\
 25 & 10 &  462827&.117 & 100 &    34 & & 45 &  4 &  819200&.272 &  50 & $-$71 & & 60 &  0 & 1085259&.811 &  20 &  $-$9 \\
 25 &  9 &  462991&.739 &  50 &  $-$7 & & 45 &  3 &  819306&.491 &  50 & $-$29 & & 61 & 10 & 1100941&.061 &  50 &    29 \\
 25 &  8 &  463139&.273 &  50 &    22 & & 45 &  2 &  819382&.243 &  80 &$-$148 & & 61 &  9 & 1101320&.232 &  50 & $-$73 \\
 25 &  7 &  463269&.567 &  50 &    28 & & 46 &  9 &  835943&.102 & 150 & $-$54 & & 61 &  6 & 1102220&.449 &  20 &     8 \\
 25 &  6 &  463382&.606 &  50 &    46 & & 46 &  6 &  836637&.725 &  80 & $-$26 & & 61 &  5 & 1102440&.856 &  30 & $-$18 \\
 25 &  5 &  463478&.295 &  50 &    26 & & 46 &  5 &  836807&.842 &  80 & $-$10 & & 61 &  3 & 1102761&.793 &  20 &     8 \\
 25 &  4 &  463556&.690 &  50 &    63 & & 46 &  4 &  836947&.184 &  80 &    67 & & 61 &  2 & 1102862&.131 &  20 &  $-$7 \\
 25 &  3 &  463617&.624 &  50 &    18 & & 46 &  3 &  837055&.463 &  50 & $-$29 & & 61 &  0 & 1102942&.430 &  20 & $-$14 \\
 25 &  2 &  463661&.238 &  50 &    58 & & 46 &  2 &  837132&.851 &  50 & $-$82 & & 62 & 12 & 1117699&.499 & 200 &   171 \\
 25 &  1 &  463687&.317 &  50 & $-$13 & & 46 &  1 &  837179&.357 &  50 & $-$52 & & 62 & 10 & 1118588&.819 &  30 &     3 \\
 25 &  0 &  463696&.065 &  50 &    17 & & 49 & 12 &  888070&.867 & 100 &     3 & & 62 &  9 & 1118973&.704 &  30 &    16 \\
 26 & 12 &  480213&.412 & 100 &    51 & & 49 & 11 &  888445&.580 & 120 &    42 & & 62 &  8 & 1119318&.436 &  20 & $-$11 \\
 26 & 11 &  480419&.811 & 100 &   143 & & 49 & 10 &  888788&.116 & 100 &  $-$1 & & 62 &  7 & 1119622&.944 &  20 & $-$18 \\
 26 &  9 &  480779&.185 &  50 & $-$16 & & 49 &  9 &  889098&.523 &  50 &    52 & & 62 &  6 & 1119887&.123 &  20 &    11 \\
 26 &  8 &  480932&.181 &  80 &$-$107 & & 49 &  6 &  889835&.075 &  50 &    23 & & 62 &  5 & 1120110&.797 &  20 &  $-$2 \\
 26 &  7 &  481067&.510 &  50 &     2 & & 49 &  5 &  890015&.474 &  50 &    40 & & 62 &  4 & 1120293&.942 &  20 &     8 \\
 26 &  6 &  481184&.797 &  50 &  $-$9 & & 49 &  4 &  890163&.134 &  50 &    17 & & 62 &  3 & 1120436&.478 &  30 &    32 \\
 26 &  5 &  481284&.187 &  50 &    49 & & 49 &  3 &  890278&.046 &  50 &     5 & & 63 & 11 & 1135800&.408 & 120 &    63 \\
 26 &  4 &  481365&.532 &  50 &    70 & & 49 &  2 &  890360&.170 &  50 &     7 & & 63 &  8 & 1136971&.561 &  30 &    17 \\
 26 &  3 &  481428&.740 &  50 &  $-$8 & & 49 &  1 &  890409&.470 &  50 &    21 & & 63 &  6 & 1137548&.452 &  20 &     5 \\
 26 &  2 &  481473&.947 &  50 & $-$23 & & 49 &  0 &  890425&.897 &  50 &    18 & & 63 &  5 & 1137775&.379 &  20 &     6 \\
 26 &  1 &  481501&.168 &  50 &    56 & & 50 & 11 &  906143&.305 & 150 & $-$87 & & 63 &  4 & 1137961&.166 &  20 &     5 \\
 26 &  0 &  481510&.165 &  50 &     6 & & 50 & 10 &  906492&.536 &  80 &    97 & & 63 &  3 & 1138105&.724 &  20 & $-$14 \\
 27 &  9 &  498564&.375 &  50 & $-$25 & & 50 &  9 &  906808&.668 &  50 &    17 & & 63 &  2 & 1138209&.052 &  20 &     5 \\
 27 &  8 &  498723&.057 & 100 &  $-$3 & & 50 &  8 &  907091&.879 &  80 & $-$30 & & 63 &  1 & 1138271&.082 &  30 &    32 \\
 27 &  7 &  498863&.213 &  80 &    12 & & 50 &  7 &  907342&.068 &  80 & $-$35 & & 64 & 11 & 1153431&.417 &  80 &    18 \\
 27 &  6 &  498984&.839 &  50 &    69 & & 50 &  6 &  907559&.150 &  50 &    15 & & 64 & 10 & 1153868&.550 &  50 &    26 \\
 27 &  5 &  499087&.724 &  50 &     8 & & 50 &  5 &  907742&.947 &  50 &    25 & & 64 &  8 & 1154619&.250 &  30 &  $-$3 \\
 27 &  3 &  499237&.616 &  50 &    24 & & 50 &  4 &  907893&.365 &  50 & $-$27 & & 64 &  7 & 1154932&.519 &  80 & $-$51 \\
 27 &  2 &  499284&.459 &  50 &  $-$1 & & 50 &  3 &  908010&.479 &  50 &  $-$7 & & 64 &  6 & 1155204&.355 &  30 &  $-$4 \\
 27 &  0 &  499322&.004 &  50 &    38 & & 51 & 12 &  923448&.206 & 100 & $-$16 & & 64 &  5 & 1155434&.482 &  30 & $-$29 \\
 28 &  9 &  516347&.309 &  80 &    49 & & 51 & 10 &  924192&.510 & 100 &  $-$0 & & 64 &  4 & 1155622&.929 &  20 & $-$12 \\
 28 &  8 &  516511&.508 &  50 &    27 & & 51 &  9 &  924514&.553 &  80 &  $-$9 & & 64 &  3 & 1155769&.523 &  80 & $-$50 \\
 28 &  7 &  516656&.554 &  50 &    18 & & 51 &  8 &  924803&.059 &  50 &    10 & & 64 &  2 & 1155874&.320 &  80 & $-$32 \\
 28 &  6 &  516782&.448 &  50 &    82 & & 51 &  7 &  925057&.871 &  50 &     9 & & 64 &  1 & 1155937&.201 &  30 & $-$34 \\
 28 &  5 &  516888&.896 &  50 & $-$26 & & 51 &  6 &  925278&.880 &  50 & $-$21 & & 64 &  0 & 1155958&.150 &  30 & $-$49 \\
 28 &  3 &  517044&.107 &  50 &    56 & & 51 &  5 &  925466&.075 &  50 &  $-$6 & & 65 &  9 & 1171901&.821 &  30 &    10 \\
 34 & 13 &  621972&.452 &  50 & $-$23 & & 51 &  4 &  925619&.334 &  50 &     5 & & 65 &  8 & 1172261&.466 &  30 & $-$24 \\
 34 & 12 &  622261&.132 &  50 &     8 & & 51 &  3 &  925738&.595 &  50 &    11 & & 65 &  7 & 1172579&.205 &  30 &    23 \\
 34 & 11 &  622527&.126 &  30 &    33 & & 51 &  2 &  925823&.803 &  50 &     3 & & 65 &  5 & 1173088&.143 &  20 &    15 \\
 34 &  9 &  622990&.550 &  50 & $-$45 & & 51 &  1 &  925874&.943 &  50 &     0 & & 65 &  3 & 1173427&.860 &  50 &  $-$7 \\
 34 &  8 &  623187&.975 &  20 &    24 & & 51 &  0 &  925891&.997 &  50 &     4 & & 65 &  2 & 1173534&.097 &  20 & $-$11 \\
 34 &  6 &  623513&.510 &  20 &    21 & & 55 & 10 &  994948&.576 & 150 & $-$48 & & 65 &  0 & 1173619&.153 &  20 &    28 \\
 34 &  5 &  623641&.561 &  20 &    17 & & 55 &  8 &  995602&.961 &  80 & $-$85 & & 66 & 12 & 1188185&.703 & 150 &$-$111 \\
 34 &  4 &  623746&.418 &  40 &    34 & & 55 &  7 &  995876&.199 &  50 &    26 & & 66 &  9 & 1189533&.633 &  80 &    73 \\
 34 &  3 &  623827&.915 &  50 & $-$56 & & 55 &  6 &  996113&.103 &  50 &     6 & & 66 &  7 & 1190220&.192 &  30 & $-$24 \\
 34 &  2 &  623886&.265 &  20 &  $-$4 & & 55 &  5 &  996313&.763 &  50 &    33 & & 66 &  6 & 1190499&.574 &  30 & $-$01 \\
 34 &  1 &  623921&.231 &  30 & $-$27 & & 55 &  3 &  996605&.769 &  50 & $-$48 & & 66 &  5 & 1190736&.074 & 100 & $-$64 \\
 34 &  0 &  623932&.881 &  30 & $-$41 & & 55 &  2 &  996697&.183 &  50 &    27 & & 66 &  4 & 1190929&.820 &  50 &     4 \\
 35 & 14 &  639388&.490 & 150 &     4 & & 55 &  1 &  996752&.051 &  50 &    77 & & 66 &  2 & 1191188&.194 &  50 & $-$36 \\
 35 & 13 &  639708&.307 & 100 & $-$38 & & 55 &  0 &  996770&.268 &  50 &    19 & & 66 &  1 & 1191252&.892 &  30 &    27 \\
 35 & 12 &  640004&.964 &  40 & $-$44 & & 56 &  9 & 1012977&.101 &  30 & $-$21 & &    &    &              &     &       \\
\end{longtable}
}
\end{center}


\begin{center}
\small{
\begin{longtable}{cr@{}lrrlcr@{}lrr}
\caption{\label{tab:CH2DCN} Quantum numbers of rotational transitions of 
CH$_2$DCN from present work, frequencies (MHz), uncertainties unc. (kHz), and 
residuals o$-$c (kHz) between observed frequencies and those calculated from 
the final set of spectroscopic parameters.} \\
\hline \hline
$J', K_a', K_c' - J'', K_a'', K_c''^a$ & \multicolumn{2}{c}{frequency} & unc. & o$-$c & &
$J', K_a', K_c' - J'', K_a'', K_c''^a$ & \multicolumn{2}{c}{frequency} & unc. & o$-$c \\
\cline{1-5} \cline{7-11}
\endfirsthead
\caption{continued.} \\
\hline \hline
$J', K_a', K_c' - J'', K_a'', K_c''^a$ & \multicolumn{2}{c}{frequency} & unc. & o$-$c & &
$J', K_a', K_c' - J'', K_a'', K_c''^a$ & \multicolumn{2}{c}{frequency} & unc. & o$-$c \\
\cline{1-5} \cline{7-11}
\endhead
\hline
\endfoot
 16,  5, d  $-$ 15,  5, d  &  277717&.406    &  15 &     1 & & 52,  2, 50 $-$ 51,  2, 49 &  903537&.321 & 100 &    79 \\
 16,  4, d  $-$ 15,  4, d  &  277761&.128    &   8 &     0 & & 52,  1, 51 $-$ 51,  1, 50 &  904004&.487 & 200 &$-$106 \\
 16,  1, 15 $-$ 15,  1, 14 &  279000&.540    &   8 &  $-$4 & & 53,  1, 53 $-$ 52,  1, 52 &  913688&.193 & 200 &$-$117 \\
 18, 10, d  $-$ 17, 10, d  &  312026&.460    &  30 & $-$33 & & 53,  5, d  $-$ 52,  5, d  &  918213&.532 & 200 &   244 \\
 18,  9, d  $-$ 17,  9, d  &  312124&.485    &  25 &  $-$1 & & 53,  2, 51 $-$ 52,  2, 50 &  920897&.991 & 200 &$-$104 \\
 18,  8, d  $-$ 17,  8, d  &  312212&.301    &  20 & $-$22 & & 53,  1, 52 $-$ 52,  1, 51 &  921264&.278 & 200 & $-$12 \\
 19,  1, 19 $-$ 18,  1, 18 &  328415&.596    &  20 &  $-$4 & & 57,  0, 57 $-$ 56,  0, 56 &  983638&.923 & 200 &   221 \\
 19, 10, d  $-$ 18, 10, d  &  329351&.951    &  30 &  $-$5 & & 57,  7, d  $-$ 56,  7, d  &  986718&.597 & 200 & $-$83 \\
 19,  3, 17 $-$ 18,  3, 16 &  329864&.188    &  20 &  $-$3 & & 57,  6, d  $-$ 56,  6, d  &  986958&.857 & 100 &    25 \\
 19,  2, 17 $-$ 18,  2, 16 &  330013&.795    &  10 &     2 & & 57,  5, d  $-$ 56,  5, d  &  987192&.298 & 100 &$-$104 \\
 28,  2, 26 $-$ 27,  2, 25 &  486412&.193    &  50 & $-$14 & & 57,  3, 54 $-$ 56,  3, 53 &  988062&.351 & 150 &    64 \\
 28,  6, d  $-$ 27,  6, d  &  485722&.367    &  50 &    55 & & 57,  1, 56 $-$ 56,  1, 55 &  990226&.073 & 150 &$-$131 \\
 28,  5, d  $-$ 27,  5, d  &  485817&.005    &  50 &    17 & & 59,  0, 59 $-$ 58,  0, 58 & 1017787&.045 & 100 & $-$42 \\
 28,  2, 27 $-$ 27,  2, 26 &  485869&.754    &  50 &    29 & & 59,  3, 56 $-$ 58,  3, 55 & 1022618&.283 & 200 &   305 \\
 28,  4, d  $-$ 27,  4, d  &  485902&.358    &  50 &     2 & & 59,  2, 57 $-$ 58,  2, 56 & 1024977&.054 & 200 &   138 \\
 28,  3, 26 $-$ 27,  3, 25 &  485983&.562    &  50 & $-$53 & & 60,  1, 60 $-$ 59,  1, 59 & 1033530&.934 & 200 & $-$76 \\
 28,  3, 25 $-$ 27,  3, 24 &  485997&.103    &  50 &    86 & & 60,  2, 59 $-$ 59,  2, 58 & 1038125&.138 & 200 &   458 \\
 28,  8, d  $-$ 27,  8, d  &  485492&.763    &  50 &    21 & & 60,  1, 59 $-$ 59,  1, 58 & 1041862&.576 & 200 &   330 \\
 28,  7, d  $-$ 27,  7, d  &  485614&.775    &  50 &    69 & & 60,  2, 58 $-$ 59,  2, 57 & 1042306&.148 & 250 &$-$269 \\
 28, 10, d  $-$ 27, 10, d  &  485203&.300    & 120 & $-$60 & & 63,  1, 63 $-$ 62,  1, 62 & 1084807&.020 & 150 & $-$95 \\
 28,  9, d  $-$ 27,  9, d  &  485355&.715    & 120 & $-$32 & & 63,  0, 63 $-$ 62,  0, 62 & 1086011&.334 & 100 & $-$55 \\
 28,  1, 27 $-$ 27,  1, 26 &  487942&.201    &  80 &     8 & & 63, 10, d  $-$ 62, 10, d  & 1089039&.445 & 200 &   160 \\
 29,  4, d  $-$ 28,  4, d  &  503235&.929    &  50 &     4 & & 63,  9, d  $-$ 62,  9, d  & 1089381&.592 & 150 &$-$123 \\
 29,  3, 27 $-$ 28,  3, 26 &  503321&.438    & 100 &  $-$2 & & 63,  2, 62 $-$ 62,  2, 61 & 1089618&.582 &  80 &    11 \\
 29,  3, 26 $-$ 28,  3, 25 &  503337&.368    & 100 & $-$40 & & 63,  8, d  $-$ 62,  8, d  & 1089695&.853 & 200 &   177 \\
 29,  2, 27 $-$ 28,  2, 26 &  503793&.683    &  50 & $-$15 & & 63,  7, d  $-$ 62,  7, d  & 1089985&.064 &  80 &    29 \\
 30,  8, d  $-$ 29,  8, d  &  520124&.625    &  80 &     0 & & 63,  6, d  $-$ 62,  6, d  & 1090257&.213 &  80 & $-$35 \\
 30,  6, d  $-$ 29,  6, d  &  520371&.360    & 100 & $-$16 & & 63,  5, d  $-$ 62,  5, d  & 1090528&.555 & 100 &     1 \\
 30,  3, 28 $-$ 29,  3, 27 &  520657&.371    & 100 &   155 & & 63,  4, 60 $-$ 62,  4, 59 & 1090822&.850 &  80 & $-$22 \\
 30,  3, 27 $-$ 29,  3, 26 &  520676&.142    & 100 &    14 & & 63,  4, 59 $-$ 62,  4, 58 & 1090859&.365 & 120 & $-$65 \\
 30,  2, 28 $-$ 29,  2, 27 &  521175&.868    & 100 & $-$24 & & 63,  3, 60 $-$ 62,  3, 59 & 1091699&.685 &  80 &    31 \\
 31,  1, 31 $-$ 30,  1, 30 &  535476&.485    & 100 &    21 & & 63,  1, 62 $-$ 62,  1, 61 & 1093420&.205 &  80 &    13 \\
 36,  1, 36 $-$ 35,  1, 35 &  621614&.941    &  30 & $-$32 & & 64, 10, d  $-$ 63, 10, d  & 1106217&.421 & 200 & $-$65 \\
 36, 11, d  $-$ 35, 11, d  &  623372&.367    & 100 &    24 & & 64,  7, d  $-$ 63,  7, d  & 1107179&.127 &  80 &    44 \\
 36,  0, 36 $-$ 35,  0, 35 &  623466&.170    &  70 & $-$63 & & 64,  1, 63 $-$ 63,  1, 62 & 1110587&.991 & 120 & $-$91 \\
 36, 10, d  $-$ 35, 10, d  &  623587&.841    &  50 & $-$40 & & 64,  2, 62 $-$ 63,  2, 61 & 1111566&.310 & 200 &   175 \\
 36,  9, d  $-$ 35,  9, d  &  623783&.719    & 100 & $-$30 & & 65,  1, 65 $-$ 64,  1, 64 & 1118961&.766 & 200 &   224 \\
 36,  7, d  $-$ 35,  7, d  &  624118&.615    &  50 &    30 & & 65,  0, 65 $-$ 64,  0, 64 & 1120088&.060 & 200 &$-$151 \\
 36,  6, d  $-$ 35,  6, d  &  624259&.658    &  70 & $-$55 & & 65,  8, d  $-$ 64,  8, d  & 1124068&.425 & 200 &   150 \\
 36,  2, 35 $-$ 35,  2, 34 &  624357&.951    &  50 &    81 & & 65,  6, d  $-$ 64,  6, d  & 1124651&.425 & 200 &    25 \\
 36,  5, d  $-$ 35,  5, d  &  624386&.762    &  70 & $-$54 & & 65,  4, 62 $-$ 64,  4, 61 & 1125246&.197 & 150 &    32 \\
 36,  3, 33 $-$ 35,  3, 32 &  624671&.021    &  50 &    56 & & 65,  3, 62 $-$ 64,  3, 61 & 1126225&.423 & 200 &   210 \\
 36,  2, 34 $-$ 35,  2, 33 &  625480&.300    &  40 & $-$32 & & 65,  1, 64 $-$ 64,  1, 63 & 1127746&.725 & 100 &    50 \\
 37,  1, 37 $-$ 36,  1, 36 &  638831&.212    &  30 & $-$51 & & 65,  2, 63 $-$ 64,  2, 62 & 1128865&.300 & 120 &    71 \\
 37, 11, d  $-$ 36, 11, d  &  640652&.376    & 100 & $-$27 & & 66,  4, 63 $-$ 65,  4, 62 & 1142450&.318 & 100 &$-$121 \\
 37, 10, d  $-$ 36, 10, d  &  640873&.836    &  70 & $-$23 & & 66,  4, 62 $-$ 65,  4, 61 & 1142500&.850 & 100 &    70 \\
 37,  9, d  $-$ 36,  9, d  &  641075&.115    &  50 & $-$46 & & 66,  3, 64 $-$ 65,  3, 63 & 1142588&.221 & 120 &    13 \\
 37,  8, d  $-$ 36,  8, d  &  641256&.800    &  50 &     4 & & 66,  1, 65 $-$ 65,  1, 64 & 1144895&.808 & 200 & $-$15 \\
 37,  5, d  $-$ 36,  5, d  &  641696&.453    &  20 &     3 & & 66,  2, 64 $-$ 65,  2, 63 & 1146157&.487 &  50 & $-$17 \\
 37,  3, 34 $-$ 36,  3, 33 &  641996&.762    &  50 & $-$64 & & 67,  1, 67 $-$ 66,  1, 66 & 1153091&.574 & 200 & $-$94 \\
 37,  1, 36 $-$ 36,  1, 35 &  644319&.793    &  40 & $-$22 & & 67, 10, d  $-$ 66, 10, d  & 1157721&.118 & 200 &    25 \\
 51,  1, 51 $-$ 50,  1, 50 &  879398&.852    & 150 & $-$74 & & 67,  7, d  $-$ 66,  7, d  & 1158730&.616 & 200 &   138 \\
 51,  6, d  $-$ 50,  6, d  &  883499&.518    & 100 &$-$149 & & 68,  1, 68 $-$ 67,  1, 67 & 1170147&.296 & 150 &$-$169 \\
 51,  5, d  $-$ 50,  5, d  &  883699&.093    & 100 &    25 & & 68,  0, 68 $-$ 67,  0, 67 & 1171160&.249 &  80 &    98 \\
 51,  3, 49 $-$ 50,  3, 48 &  884076&.317    & 150 &$-$113 & & 68,  8, d  $-$ 67,  8, d  & 1175587&.939 & 200 &$-$178 \\
 51,  2, 49 $-$ 50,  2, 48 &  886173&.147    & 200 &   119 & & 68,  6, d  $-$ 67,  6, d  & 1176204&.005 & 200 &$-$226 \\
 52,  1, 52 $-$ 51,  1, 51 &  896546&.406    & 150 &   179 & & 68,  1, 67 $-$ 67,  1, 66 & 1179165&.037 & 200 &$-$191 \\
 52,  0, 52 $-$ 51,  0, 51 &  898157&.427    & 150 &    45 & & 69,  1, 69 $-$ 68,  1, 68 & 1187196&.902 & 200 &$-$100 \\
 52,  2, 51 $-$ 51,  2, 50 &  900538&.159$^b$& 150 &  $-$1 & & 69,  0, 69 $-$ 68,  0, 68 & 1188172&.743 & 200 &    74 \\
 52,  7, d  $-$ 51,  7, d  &  900538&.159$^b$& 150 &  $-$1 & & 69,  1, 68 $-$ 68,  1, 67 & 1196285&.083 & 200 &$-$134 \\
\end{longtable}
}
\end{center}
\footnotesize{
$^a$ A ''d'' given for $K_c$ signals that the asymmetry doubling has not been resolved for the respective $K_a$.
The uncertainty and the residual refer to the average. Since $J - K_a - K_c = 0$ or 1, $K_c$ is redundant in such case.\\
$^b$ Blended lines treated as intensity weighted average in the fit.\\
}


\begin{center}
\small{
\begin{longtable}{rrr@{}lrrlrrr@{}lrrlrrr@{}lrr}
\caption{\label{tab:2x13} Lower state quantum numbers of rotational transitions 
of $^{13}$CH$_3^{13}$CN, frequencies (MHz), uncertainties unc. (kHz), and 
residuals o$-$c (kHz) between observed frequencies and those calculated from 
the final set of spectroscopic parameters.} \\
\hline \hline
$J''$ & $K''$ & \multicolumn{2}{c}{frequency} & unc. & o$-$c & &
$J''$ & $K''$ & \multicolumn{2}{c}{frequency} & unc. & o$-$c & &
$J''$ & $K''$ & \multicolumn{2}{c}{frequency} & unc. & o$-$c \\
\cline{1-6} \cline{8-13} \cline{15-20}
\endfirsthead
\caption{continued.} \\
\hline \hline
$J''$ & $K''$ & \multicolumn{2}{c}{frequency} & unc. & o$-$c & &
$J''$ & $K''$ & \multicolumn{2}{c}{frequency} & unc. & o$-$c & &
$J''$ & $K''$ & \multicolumn{2}{c}{frequency} & unc. & o$-$c \\
\cline{1-6} \cline{8-13} \cline{15-20}
\endhead
\hline
\endfoot
 13 & 4 &  249849&.289 &  30 &    30 & & 18 & 5 &  338978&.878 &  50 &     2 & & 51 & 2 &  926328&.788 & 200 &$-$108 \\
 13 & 3 &  249881&.971 &  10 &  $-$5 & & 18 & 2 &  339111&.819 &  15 &  $-$6 & & 51 & 1 &  926379&.428 & 200 &$-$194 \\
 14 & 4 &  267689&.340 &  20 &  $-$8 & & 25 & 3 &  463885&.870 &  50 &    22 & & 54 & 3 &  979425&.780 & 150 &    20 \\
 14 & 3 &  267724&.401 &  10 &     9 & & 25 & 2 &  463929&.028 &  50 & $-$11 & & 60 & 6 & 1085127&.880 & 150 &    34 \\
 14 & 2 &  267749&.425 &  10 &  $-$7 & & 27 & 3 &  499525&.679 & 100 &    88 & & 60 & 4 & 1085519&.654 & 150 & $-$68 \\
 14 & 1 &  267764&.473 &  15 &    12 & & 34 & 7 &  623721&.761 & 200 & $-$22 & & 60 & 3 & 1085656&.847 & 200 &$-$150 \\
 15 & 6 &  285421&.490 &  40 &    21 & & 34 & 5 &  623998&.639 & 150 & $-$25 & & 60 & 0 & 1085833&.407 & 200 &$-$177 \\
 15 & 4 &  285528&.155 &  30 &    18 & & 34 & 4 &  624102&.519 & 150 & $-$85 & & 61 & 9 & 1101913&.821 & 100 & $-$11 \\
 15 & 3 &  285565&.500 &  15 &  $-$4 & & 34 & 3 &  624183&.422 & 100 & $-$68 & & 61 & 5 & 1103025&.791 & 150 &    69 \\
 15 & 2 &  285592&.188 &  15 & $-$17 & & 34 & 2 &  624241&.332 &  50 &    45 & & 61 & 4 & 1103204&.664 & 150 &$-$125 \\
 15 & 1 &  285608&.237 &  30 &     7 & & 35 & 9 &  641127&.171 & 200 &   161 & & 61 & 2 & 1103443&.760 & 150 &    55 \\
 16 & 6 &  303252&.220 &  25 & $-$24 & & 35 & 3 &  641980&.261 &  80 & $-$29 & & 61 & 1 & 1103503&.550 & 150 &    88 \\
 16 & 1 &  303450&.598 &  15 &  $-$8 & & 35 & 1 &  642075&.314 &  80 & $-$31 & & 62 & 6 & 1120480&.948 & 200 &   154 \\
 17 & 6 &  321081&.547 &  15 &  $-$6 & & 50 & 3 &  908507&.997 & 150 & $-$88 & & 62 & 4 & 1120884&.433 & 200 & $-$34 \\
 17 & 5 &  321147&.495 &  20 &     9 & & 50 & 2 &  908591&.195 & 150 &   123 & & 62 & 3 & 1121025&.872 & 100 &  $-$3 \\
 17 & 4 &  321201&.474 &  15 &     8 & & 50 & 0 &  908657&.612 & 150 &   132 & & 62 & 2 & 1121126&.880 & 150 & $-$39 \\
 17 & 3 &  321243&.472 &   5 &  $-$0 & & 51 & 6 &  925788&.347 & 200 & $-$80 & & 63 & 2 & 1138804&.705 & 200 &    60 \\
 17 & 2 &  321273&.504 &  10 &    15 & & 51 & 3 &  926244&.313 & 200 & $-$61 & & 63 & 1 & 1138866&.331 & 200 &   163 \\
 17 & 1 &  321291&.490 &  15 & $-$13 & &    &   &        &     &     &       & &    &   &        &     &     &       \\
\end{longtable}
}
\end{center}


\begin{thebibliography}{}

\bibitem[Ag{\'u}ndez et al.(2008)]{MeCN_etc_IRC+10216} 
Ag{\'u}ndez, M., Fonfr{\'{\i}}a, J.~P., Cernicharo, J., et al. 
2008, \aap, 479, 493 

\bibitem[Anttila et al.(1993)]{MeCN-A} 
Anttila, R., Horneman, V.-M., Koivusaari, M. \& Paso, R. 
1993, J. Mol. Spectrosc., 157, 198

\bibitem[Bauer et al.(1975)]{MeCN-15N} 
Bauer, A., Tarrago, G., \& Remy, A. 
1975, J. Mol. Spectrosc., 58, 111 

\bibitem[Belloche et al.(2009)]{det-PrCN_EtFo} 
Belloche, A., Garrod, R.~T., M{\"u}ller, H.~S.~P., et al. 
2009, \aap, 499, 215

\bibitem[Boucher et al.(1977)]{MeCN-Lille} 
Boucher, D., Burie, J., Demaison, J., et al.
1977, J. Mol. Spectrosc., 64, 290 

\bibitem[Cazaux et al.(2003)]{MeCN_IRAS16293} 
Cazaux, S., Tielens, A.~G.~G.~M., Ceccarelli, C., et al.
2003, \apjl, 593, L51 

\bibitem[Cazzoli \& Puzzarini(2006)]{MeCN-LD} 
Cazzoli, G., \& Puzzarini, C.\ 
2006, J. Mol. Spectrosc., 240, 153; 
Erratum, 
2008, J. Mol. Spectrosc., 247, 187 

\bibitem[Cummins et al.(1983)]{MeCN-T_kin} 
Cummins, S.~E., Green, S., Thaddeus, P., \& Linke, R.~A.\ 
1983, \apj, 266, 331 

\bibitem[Demaison et al.(1979)]{MeCN-isos_1979} 
Demaison, J., Dubrulle, A., Boucher, D., 
1979, J. Mol. Spectrosc., 76, 1

\bibitem[Drouin et al.(2005)]{JPL-spectrometer} 
Drouin, B.~J., Maiwald, F.~W., \& Pearson, J.~C. 
2005, Rev. Sci. Instr., 76, Art.-No. 093113

\bibitem[Gadhi et al.(1979)]{MeCN-dipole}
Gadhi, J., Lahrouni, A., Legrand, J., \& Demaison, J. 
1995, J. Chem. Phys., 92, 1984

\bibitem[Gerin et al.(1992)]{det-CH2DCN} 
Gerin, M., Combes, F., Wlodarczak, et al.
1992, \aap, 259, L35 

\bibitem[Haekel \& M\"ader(1989)]{MeCN-15_1-0} 
Haekel, J. \& M\"ader, H.
1989, J. Quant. Spectrosc. Radiat. Transfer, 41, 9

\bibitem[Halonen \& Mills(1978)]{CHD2CN-rot} 
Halonen, L., \& Mills, I.~M.\ 
1978, J. Mol. Spectrosc., 73, 494

\bibitem[Kukolich et al.(1973)]{MeCN_1-0} 
Kukolich, S.~G., Ruben, D.~J., Wang, J.~H.~S., \& Williams, J.~R. 
1973, J. Chem. Phys., 58, 3155

\bibitem[Kukolich (1982)]{MeCN-12-13b_2-1} 
Kukolich, S.~G., 
1982, J. Chem. Phys., 76, 97

\bibitem[Le Guennec et al.(1992)]{CH2DCN-rot} 
Le Guennec, M., Wlodarczak, G., Burie, J., \& Demaison, J.\ 
1992, J. Mol. Spectrosc., 154, 305 

\bibitem[Matthews \& Sears(1983)]{MeCN-TMC-1} 
Matthews, H.~E., \& Sears, T.~J.\ 
1983, \apjl, 267, L53 

\bibitem[Mauersberger et al.(1991)]{MeCN_MeCCH_extragal} 
Mauersberger, R., Henkel, C., Walmsley, C.~M., et al.
1991, \aap, 247, 307 

\bibitem[Mito et al.(1984)]{MeCN-15-dipole} 
Mito, A., Sakai, J., \& Katayama, M. 
1984, J. Mol. Spectrosc., 103, 26

\bibitem[M{\"u}ller et al.(2000)]{SO-17O} 
M{\"u}ller, H.~S.~P., Farhoomand, J., Cohen, E.~A., et al. 
2000, J. Mol. Spectrosc., 201, 1 

\bibitem[M{\"u}ller et al.(2001)]{CDMS_1}
M{\"u}ller, H.~S.~P., Thorwirth, S., Roth, D.~A.,
\& Winnewisser, G.
2001, A\&A, 370, L49

\bibitem[M{\"u}ller et al.(2005)]{CDMS_2}
M{\"u}ller, H.~S.~P., Schl{\"o}der, F., Stutzki, J.,
\& Winnewisser, G.
2005, J. Mol. Struct, 742, 215

\bibitem[M{\"u}ller \& Br{\"u}nken(2005)]{SO2_2005} 
M{\"u}ller, H.~S.~P., \& Br{\"u}nken, S. 
2005, J. Mol. Spectrosc., 232, 213 

\bibitem[M{\"u}ller et al.(2007a)]{SiS_2007}
M{\"u}ller, H.~S.~P., McCarthy, M. C., Bizzocchi, L., et al.
2007a, Phys. Chem. Chem. Phys., 9, 1579

\bibitem[M{\"u}ller et al.(2007b)]{MeCN-OSU}
M\"uller, H.~S.~P., Drouin, B. J., Pearson, J. C., et al.
2007b, contribution WG03, presented at the 62nd International 
Symposium on Molecular Spectroscopy, June 18 – 22, 2007, 
Columbus, Ohio, USA; see also 
http://molspect.chemistry.ohio-state.edu/symposium\_62/symposium/Abstracts/p074.pdf

\bibitem[M{\"u}ller et al.(2008)]{13C-VyCN_2008} 
M{\"u}ller, H.~S.~P., Belloche, A., Menten, K.~M., et. al.
2008, J. Mol. Spectrosc., 251, 319 

\bibitem[Nummelin et al.(1998)]{Sgr-B2_Nummelin} 
Nummelin, A., Bergman, P., Hjalmarson, \r{A}., et al.
1998, \apjs, 117, 427

\bibitem[Pavone et al.(1990)]{MeCN-TuFIR} 
Pavone, F.~S., Zink, L.~R., Prevedelli, M., et al.
1990, J. Mol. Spectrosc., 144, 45

\bibitem[Pearson \& M\"uller(1996)]{MeCN_rot-isos_1996} 
Pearson, J.~C., \& M\"uller, H.~S.~P.\ 
1996, \apj, 471, 1067 

\bibitem[Pickett et al.(1998)]{JPL-catalog} 
Pickett, H.~M., Poynter, R.~L., Cohen, E.~A., et al. 
1998, J. Quant.  Spectrosc. Radiat. Transfer, 60, 883

\bibitem[Ring et al.(1947)]{MeCN_1st-MW} 
Ring, H., Edwards, H., Kessler, M., \& Gordy, W. 
1947, Phys. Rev., 72, 1262

\bibitem[{\v S}ime{\v c}kov{\'a} et al.(2004)]{MeCN-v08} 
{\v S}ime{\v c}kov{\'a}, M., Urban, {\v S}., Fuchs, U., et al. 
2004, J. Mol. Spectrosc., 226, 123 

\bibitem[Solomon et al.(1971)]{det-MeCN} 
Solomon, P.~M., Jefferts, K.~B., Penzias, A.~A., \& Wilson, R.~W.\ 
1971, \apjl, 168, L107 

\bibitem[Sutton et al.(1985)]{Orion-survey_13CH3CN} 
Sutton, E.~C., Blake, G.~A., Masson, C.~R., \& Phillips, T.~G. 
1985, \apjs, 58, 341 

\bibitem[Tam et al.(1988)]{MeCN-2x13} 
Tam, H., An, I., \& Roberts, J. A.
1988, J. Mol. Spectrosc., 129, 202

\bibitem[Thomas et al.(1955)]{CH2DCN_2-1} 
Thomas, L.~F., Sherrard, E.~J. \& Sheridan, J.
1955, Trans. Faraday Soc., 51, 619 

\bibitem[Ulich \& Conklin(1974)]{MeCN-Kohoutek} 
Ulich, B.~L., \& Conklin, E.~K.\ 
1974, \nat, 248, 121 

\bibitem[Wilson \& Rood(1994)]{isotopic_abundances} 
Wilson, T.~L., \& Rood, R. 
1994, \araa, 32, 191

\bibitem[Winnewisser et al.(1994)]{BWO-THz_spec}
Winnewisser, G., Krupnov, A. F., Tretyakov, M. Y., et al.
1994, J. Mol. Spectrosc., 165, 294

\end{thebibliography}
\end{document}